\documentclass[10pt,journal,compsoc]{IEEEtran}

\usepackage{graphicx}
\usepackage[caption=false]{subfig}
\usepackage{url}
\usepackage{amsmath,amsthm}
\usepackage{mathtools}
\usepackage{bm}
\usepackage{lscape}
\usepackage{acronym}
\usepackage{color}
\usepackage{flushend}
\usepackage{float}
\usepackage{dblfloatfix}
\usepackage{color}

\newcommand{\beq}{\begin{equation}}
\newcommand{\eeq}{\end{equation}}
\newcommand{\beqa}{\begin{eqnarray}}
\newcommand{\eeqa}{\end{eqnarray}}
\DeclareMathOperator*{\argmin}{arg\,min}

\definecolor{newgreen}{rgb}{0.133,0.545,0.133}

\begin{document}

\title{Availability Evaluation of Multi-tenant Service Function Chaining Infrastructures by Multidimensional Universal Generating Function}

\author{Mario~Di~Mauro,~\IEEEmembership{Member,~IEEE,}
	Maurizio~Longo,~\IEEEmembership{Member,~IEEE,}
        Fabio~Postiglione
\IEEEcompsocitemizethanks{\IEEEcompsocthanksitem M. Di Mauro, M. Longo, F. Postiglione are with the Department of Information and Electrical Engineering and Applied Mathematics (DIEM), University of Salerno, 84084, Fisciano, Italy.
\protect\\
E-mail: \{mdimauro,longo,fpostiglione\}@unisa.it
}
}

\IEEEtitleabstractindextext{%
\begin{abstract}
The Network Function Virtualization (NFV) paradigm has been devised as an enabler of next generation network infrastructures by speeding up the provisioning and the composition of novel network services. The latter are implemented via a chain of virtualized network functions, a process known as Service Function Chaining. In this paper, we evaluate the availability of multi-tenant SFC infrastructures, where every network function is modeled as a multi-state system and is shared among different and independent tenants. To this aim, we propose a Universal Generating Function (UGF) approach, suitably extended to handle performance vectors, that we call Multidimensional UGF.
This novel methodology is validated in a realistic multi-tenant telecommunication network scenario, where the service chain is composed by the network elements of an IP Multimedia Subsystem implemented via NFV. A steady-state availability evaluation of such an exemplary system is presented and a redundancy optimization problem is solved, so providing the SFC infrastructure which minimizes deployment cost while respecting a given availability requirement.
\end{abstract}

}

\maketitle

\IEEEdisplaynontitleabstractindextext

\IEEEpeerreviewmaketitle

\IEEEraisesectionheading{\section{Introduction}\label{sec:intro}}
\IEEEPARstart{I}n the era of fifth generation (5G) telecommunication systems, the design, management and deployment of complex architectures have dramatically boosted, due to the increasing demand of network resources by more and more connected devices as smartphones, laptops, tablets, sensor networks and other kinds of smart objects.
Facing these issues, several telecom operators have established an industry specifications group  providing guidelines for Network Functions Virtualization (NFV) \cite{etsi2012}. 
NFV is the network concept aiming to virtualize the whole class of network node functions (routers, firewalls, load balancers and others) into building blocks that may be interconnected to create communication services. 
The resulting architecture includes a set of Virtualized Network Functions (VNFs), conveniently arranged to create innovative network services or to define new \textit{as-a-service} models such as VNFaaS \cite{cotroneo17}, often in conjunction with Software Defined Networking (SDN) \cite{McKeown2008} aimed at controlling the composition logic \cite{matias2015} and governing some security mechanisms \cite{ali2015}. 
New services can be designed by means of the so-called Service Function Chaining (SFC) process, which consists in selecting specific VNFs to be connected and traversed in a predefined order \cite{sfc2015}, \cite{intel2016}.

In this paper, we present an availability analysis of an SFC infrastructure guaranteeing the so-called ``five nines" availability requirement (no more than 5 minutes and 26 seconds system downtime per year) as invoked in typical Service Level Agreements (SLAs). 
In particular, we focus on a multi-tenant SFC architecture, where several operators (\textit{aka} tenants) share the existing VNFs to provide specific services. Such VNFs are prone to failure (and consequent repair) activities, that could alter in many ways the overall SFC functioning. For instance, a VNF under repair and, hence, temporarily out-of-service, might have a huge impact on one tenant but minimal influence on another one. 

Our analysis exploits the remarkable properties of the Universal Generating Function (UGF), a formalism originally introduced in \cite{ushakov1986}. In particular, we propose an  extended version that we call Multidimensional UGF (MUGF), useful to handle multidimensional performance figures. 

The paper is organized as follows. In Section \ref{sec:Motiv} we advance a general perspective of the considered problem. Section \ref{sec:RW} presents a review of the most significant related work. Some details about the SFC paradigm, along with a brief description of NFV architecture, are given in Section \ref{sec:SFC}. In Section \ref{sec:perfmetric} we outline a multi-state performance model of an SFC architecture, which accounts for failures and repair actions. 
In Section \ref{sec:ugf} we study the steady-state availability of the system by introducing the novel MUGF approach. 
In Section \ref{sec:results} we develop the solution of a redundancy optimization problem for an exemplary SFC infrastructure representing a multi-tenant 5G telecommunication system. Finally, Section \ref{sec:conclusions} draws the main conclusions and indicates further research prospects.

\section{Motivations and Problem Statement}\label{sec:Motiv}

Network operators are moving towards NFV-based infrastructures to efficiently reduce deployment efforts and to expedite the provisioning of new services. 
Perfectly inserted in an NFV ecosystem, SFC refers to a technique for selecting "network elements" to be traversed in a predefined order to provide a specific service. 
Consequently, IP packets of a data flow are processed in a sequential manner by a series of network service functions (e.g., network address translator, load balancer, firewall, deep packet inspector) that are implemented as VNFs.

An exemplary use case is offered in the present work, where the SFC is composed by virtualized nodes of an IP Multimedia Subsystem (IMS), a key infrastructure deployed in the core networks of next generation telecommunication systems. 
In this regard, we want to highlight that, being mainly focused on an availability problem, we consider, for the sake of simplicity, a high-level perspective of the IMS service chain, as often contemplated in the technical literature on SFC infrastructures (e.g., \cite{sendi16}, \cite{sun18}).

One of the most valuable advantages provided by SFCs in terms of cost reduction concerns the possibility to share network functionalities among different operators.
Such is the case of a {\em multi-tenant} SFC infrastructure, where VNF resources are (not necessarily equally) allocated for each operator, and hence posing a resource sharing problem \cite{mehraghdam2014}. This problem also arises in technological scenarios, as described in guidelines proposed by NEC Corporation \cite{nec2014}, where Long Term Evolution (LTE) telecommunication nodes are shared (once virtualized) among different mobile providers.

In real operations, VNFs are affected by hardware and software faults that reduce resources allocated to one or more tenants, and, typically, some repair actions are executed. Consequently, performance levels exhibited by a multi-tenant SFC vary along time and can differ from one operator to another.

For each single tenant, an SFC is considered available when it guarantees a given performance level to that tenant, a condition depending on usable resources.  
In the presence of faults, it is crucial to conceive some redundancy methods, that, obviously, entail a trade-off between costs and availability targets. 

The availability analysis of a system characterized by different performance levels can benefit of a Multi-State System (MSS) representation \cite{Levitin2003}.
In particular, the availability of MSS complex systems, composed in turn by different MSS subsystems, can be faced by the UGF approach. Such an approach allows to characterize the overall system performance distribution, and thus its availability, by composing the performance distributions of its subsystems via some appropriate operators, in a computationally efficient way \cite{Levitin2005}. However, it can only be used to assess one-dimensional performance characteristics.
On the other hand, in a multi-tenant SFC scenario, the performance levels of the operators vary from one operator to the other, thus, multivariate performance characteristics must be considered to evaluate the entire system availability. 

In the present work, we address all of the aforementioned issues by offering three original contributions: \textit{i)} we model a multi-tenant SFC infrastructure as an MSS by conveniently combining some composition operators; \textit{ii)} we propose an extended version of the UGF technique, referred to as Multidimensional UGF (MUGF), dealing with performance vectors, and then applicable to complex scenarios such as those represented by multi-tenant network architectures; \textit{iii)} we perform an availability analysis and solve a redundancy optimization problem in a realistic scenario of a virtualized 5G telecommunication infrastructure, as a profitable example of a multi-tenant SFC. Finally, a sensitivity analysis is carried out to assess the robustness of the considered system with respect to variation of critical parameters value.

\section{Related work}
\label{sec:RW}

In the last years, the scientific community has devoted an increasing interest to the issue of availability assessment of novel cloud-oriented architectures \cite{Ghosh2014}. In this section, without pretence to be exhaustive, we present a review of recent papers that have addressed problems affine to ours.
In many cases, the availability problems in cloud infrastructures are solved by proposing algorithms based on optimal allocations of virtual backup resources in order to prevent possible faults of main elements, but without considering (or partially considering) a failure/repair model. The authors in \cite{fan2017}, for example, analyze the availability problem (with regards to the minimum number of off-site backup VNFs to be deployed) of an SFC whose model includes only failures actions but not repair activities.
Similarly, in \cite{kong2017} the problem of distributing VNF replicas between the primary and backup paths to maximize the SFC's availability has been tackled through a heuristic algorithm where a failure/repair model is not addressed. The work in \cite{alameddine2017}, although not considering the chaining structure of VNFs, examines the problem of providing service availability with bandwidth guarantees through the deployment of redundant virtual machines (VMs) in a multi-tenant environment. The focus there is on the design of a protection plan where each backup VM should protect one or more primary VMs in case of failure, and no VM failure/repair model is included, being the design of the optimal backup infrastructure out of the scope.

Even when a failure/repair model is present, the availability evaluation typically encompasses non-MSS models as in \cite{sebastio18}, where an availability analysis of container-based architectures is carried on by applying some non-state-space and state-space models. Similarly, the authors of \cite{sousa14} perform an availability analysis of cloud infrastructures by exploiting the Stochastic Petri Net framework aimed at finding optimal redundancy, considering  non-MSS models of failure/repair activities. 

Approaches exploiting the UGF function (to handle multi-state system representations) are adopted in \cite{globecom10} and \cite{Guida2013}, where Markov and semi-Markov models are used, respectively, to characterize a single tenant IMS architecture.
Another UGF-based method is employed to model physical and virtual machine failures in \cite{Sun2016}, where a single tenant cloud-based environment is implicitly assumed.

Consequently, trying to fill in the gap in existing literature, in this paper we describe the novel MUGF method (previously just sketched in \cite{dimauro2017}) in order to enclose in an unified framework the ability of handling complex systems modeled by MSSs (as VNFs in a multi-tenant environment), and the capacity to evaluate, through series/parallel MUGF operators, the availability of a \textit{chained} system (such as an SFC infrastructure).

\section{Service Function Chaining in an NFV Environment}
\label{sec:SFC}

In this Section we introduce the NFV architecture and detail a multi-tenant SFC infrastructure in line with the current standardization process.

\subsection{The NFV architecture}

\begin{figure}[t!]
	\centering
	\captionsetup{justification=centering}
	\includegraphics[scale=0.4]{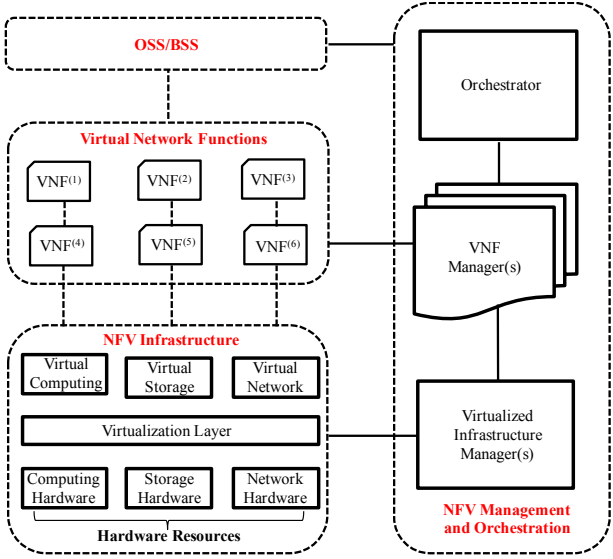}
	\caption{The reference framework of the NFV architecture.}
	\label{nfv_arch}
\end{figure}

NFV solutions offer such benefits as: \emph{i) cost saving}, resulting from the use of generic and cheaper hardware platforms rather than dedicated and costly ones;   \emph{ii) scalability}, meant as the possibility of increasing or reducing the utilization of network equipments; \emph{iii) flexibility}, achieved by faster deployment procedures of new services; \emph{iv) security}, enforced by separation and isolation of network functions.
In the NFV architectural framework the functional blocks are assorted in three domains, as per Fig. \ref{nfv_arch}:
\begin{itemize}
	\item \textit{Virtual Network Functions}, containing all the VNFs, namely the virtualized instances of network functions (routers, firewalls, load balancers, etc.), replacing traditional appliances; 
	\item \textit{NFV Infrastructure}, containing all the hardware and software components useful to build VNFs, possibly distributed across several locations; 
	\item \textit{NFV management and orchestration}, containing the modules in charge of managing the VNFs functions, e.g. allocation of computing resources, storage and network connectivity to VNFs, root cause analysis, collection of fault information. 
\end{itemize}

Such domains are supervised by the Operating and Business Support Systems (OSS/BSS) in order to guarantee network performances, customer support and charging/billing operations.

{Being interested in a multi-tenant SFC model, we specifically focus on VNFs for multi-operator scenarios and on the NFV infrastructure domain, according to the framework in Fig. \ref{nfv_arch}, where OSS/BSS and NFV Management and Orchestration domains are not considered for our purposes.} 

\subsection{A generic multi-tenant SFC model}

Some providers are experimenting the deployment of multi-tenant SFCs where virtual resources are shared among different service operators. {Three} remarkable examples are reported below.

A commercial solution of a service chain designed for the LTE mobile telecommunication world has been presented in \cite{nec2014}, where information flows of different telecom operators traverse common virtualized nodes. Some specifications about the virtualization (i.e. the deployment as VNFs chain) of main LTE nodes are also provided, such as Mobile Management Entity (vMME), Serving and Packet Data Network Gateway (vS/P-GW), Home Subscriber Server (vHSS), Policy Control and Charging Rules Function (vPCRF).

{An exemplary implementation of a multi-tenant telecommunication system is offered by the Gateway Core Network (GWCN) \cite{etsi123251,nec2013} for mobile networks, wherein more than one tenant share a consistent part of the underlying network infrastructure, so that, a tradeoff between security needs and cost savings arises. Obviously, the GWCN design has to be carefully planned by guaranteeing a satisfactory degree of isolation between tenants for security reasons (indeed, GWCN is often implemented among operators having strong commercial agreements).}

Another example (inspiring the experiment in Section \ref{sec:results}) is given by the novel 5G telecommunication network scenario in \cite{eri2014}, proposing a virtualized IP Multimedia Subsystem architecture composed by virtualized Call Session Control Function (CSCF) nodes shared among different service providers.

\begin{figure}[t!]
	\centering
	\captionsetup{justification=centering}
	\includegraphics[scale=0.3, angle=90]{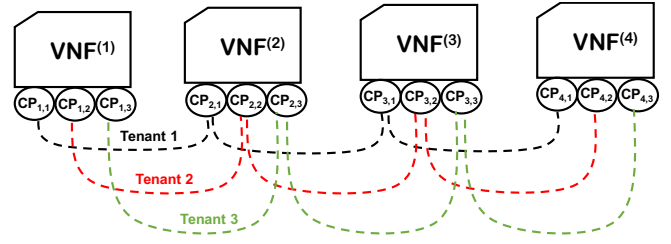}
	\caption{A multi-tenant SFC architecture with $4$ VNFs shared among $3$ tenants, each one accessing dedicated VNF instances via Connection Points (CPs). CP$_{m,i}$ refers to the connection point of VNF$^{(m)}$ traversed by tenant $i$, $i=1,\dots,K$.} 
	\label{fig:vnffg}
\end{figure}

Stemming from the analysis of the above examples, it is possible to derive a useful abstraction of a multi-tenant SFC infrastructure, that can be modeled as a chain of VNFs accessed by different operators through dedicated interfaces called Connection Points (CPs). 

One such multi-tenant SFC system is shown in Fig. \ref{fig:vnffg} where four VNFs are shared among three tenants. Each tenant $i$ ($i=1,\dots,K$) has its own dedicated access to the VNF$^{(m)}$ via connection point CP$_{m,i}$.

It is worth noting that, in this example, all tenants exploit the {\it same} SFC (the $4$ VNFs traversed in the same order), but, in principle, they could also share a subset of VNFs or simply traverse them in a diverse order, resulting in {\it different} SFCs.  {In the latter case (not considered in this work), a classifier dispatching different flows to corresponding SFCs might be needed, and it should be added at the beginning of the SFC model.} 

\section{Availability analysis of a Service Chain based on an NFV infrastructure}
\label{sec:perfmetric}

We recall that an MSS is characterized by a finite number of states representing as many performance levels.
For instance, a binary system is the simplest MSS with only two different states: perfect working and total failure.

An SFC can be regarded as an MSS, where the component VNFs are MSS subsystems as well. Firstly, we propose a performance model for a single VNF and then we provide an approach to evaluate the availability of a multi-tenant SFC system.

\subsection {A VNF multi-state performance model}
\label{sec:mss}

\begin{figure}[t]
	\centering
	\captionsetup{justification=centering}
	\includegraphics[scale=0.31, angle=90]{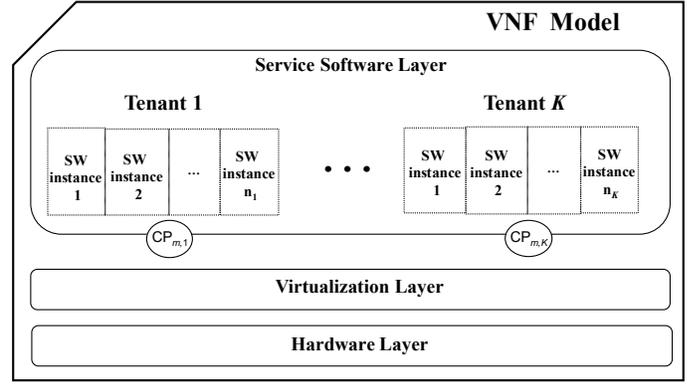}
	\caption{Schematic representation of VNF hosting $K$ tenants. The service software layer is represented by a set of tenants, each one managing some instances. The virtualization layer represents the interface with underlying physical resources {(hardware layer)}.}
	\label{vnf_layout}
\end{figure}

In Fig. \ref{vnf_layout}, we sketch a multi-state performance model of a single VNF of a multi-tenant SFC system serving $K$ different operators.
We consider a typical performance metric for telecommunication service operators, namely the number of service requests that the VNF segment devoted to a single tenant is able to manage, typically referred to as \textit{serving capacity}. The proposed approach, however, can be easily extended to other performance metrics of interest.

From an availability point of view, we propose to model a single VNF as composed by {the following three layers} (see Fig. \ref{vnf_layout}): 

\begin{itemize}
	\item a \textit{service software layer}, representing some (identical) software instances that implement the VNF serving logic and work in parallel. Each instance is modeled as a component with two states: ``active" (i.e. perfect functioning) and ``failed" (i.e.complete failure). An instance has a serving capacity equal to $\gamma$ when active, and to $0$ if failed. Tenant $i$ is supposed to manage $n_i$ software instances and to balance the load among them, where $n_i$ is selected according to some performance and availability requirements ($i=1,...,K$). For example, if a single instance in a virtualized LTE signaling node can manage up to $1000$ requests, a tenant with a performance requirement of $5800$ requests needs $6$ software instances at least;
	\item a \textit{virtualization layer}, also known as \textit{hypervisor}, representing an element able to manage communications between hardware resources and software modules for each tenant $i$, and accessed via the corresponding CP. It is modeled as a two-state model: ``active" or ``failed";
	\item {a \textit{hardware layer}, embodying hardware resources in the NFV Infrastructure domain (see Fig. \ref{nfv_arch}). Similar to the virtualization layer, a two-state model is assumed.} 
\end{itemize}


Furthermore, we suppose that: \textit{i)} software, virtualization layer and {hardware} failures are independent Homogeneous Poisson Processes (HPPs) with rates $\lambda_{s_i}$ (for tenant $i=1,...,K$), $\lambda_{v}$ and {$\lambda_{h}$}, respectively; \textit{ii)} software, virtualization layer and {hardware} repair times are independent and exponential random variables with rates $\mu_{s_i}$  ($i=1,...,K$), $\mu_{v}$ and {$\mu_{h}$}, respectively.

By arranging in each state all the numbers of active software instances managed by the tenants, the resulting VNF multi-state model is the homogeneous Continuous-Time Markov Chain (CTMC) depicted in Fig. \ref{mega_seph}, where:

\begin{figure*}[t]
	\centering
	\captionsetup{justification=centering}
	\includegraphics[scale=0.6, angle=90]{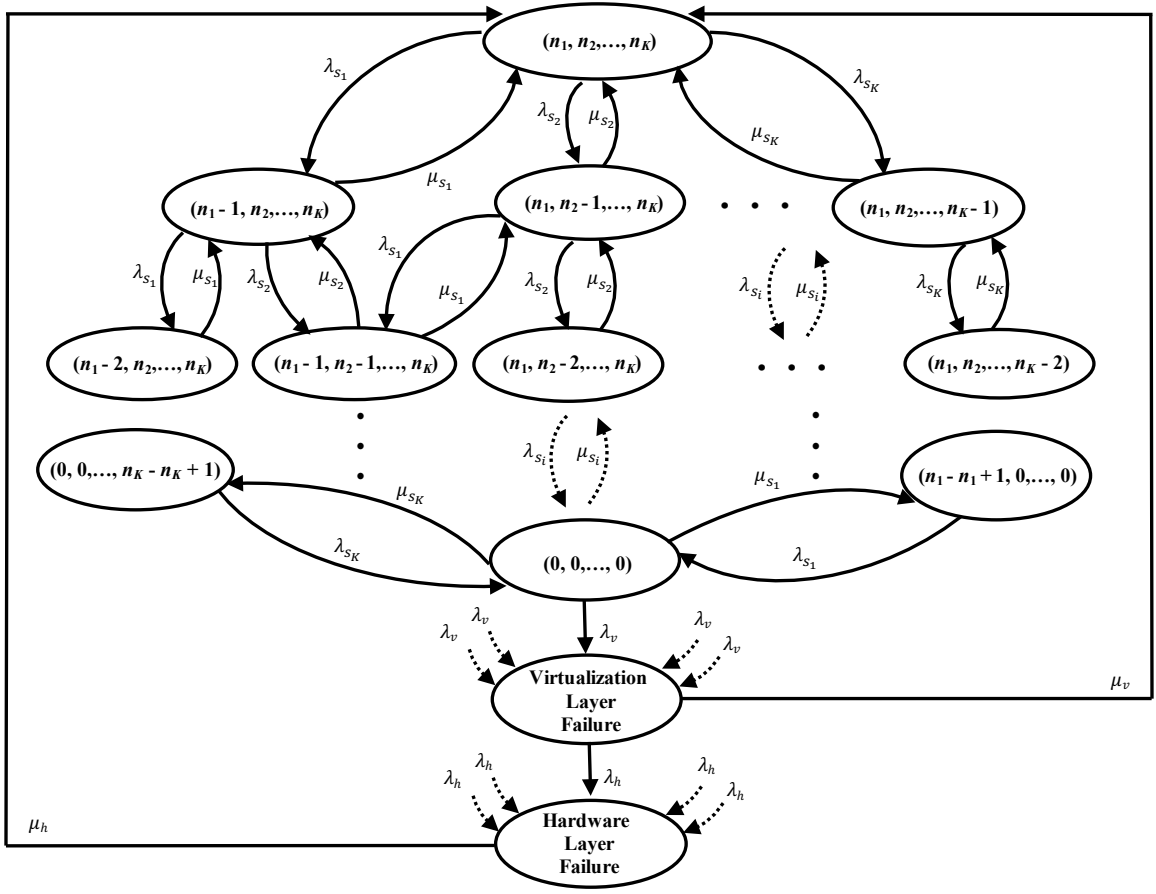}
	\caption{Multi-state model of a generic VNF. Each state contains the number of active software instances for all $K$ tenants.}
	\label{mega_seph}
\end{figure*} 

\begin{itemize}
	\item A generic state $j$ is described by the $K$-dimensional vector \boldsymbol{$\alpha$}$_j = $ $(\alpha_{1,j},...,\alpha_{K,j})$, where $\alpha_{i,j}\in{0,1,...,n_i}$ is the number of active software instances managed by tenant $i$ in that state; the initial state vector $(n_1,...,n_K)$ refers to a fully working system where the maximum number of active software instances for each tenant is available. For example, the vector $(n_1,...,n_i - 1,...,n_K)$ refers to a state where a single software instance devoted to tenant $i$ is down. 
	\item The $K$-dimensional state vector $(0,0,...,0)$ refers to a state where no software instance is active, whether the hypervisor is working or not.
	\item The state \textit{Virtualization Layer Failure} (VLF) refers to the hypervisor failure condition causing the crash of {all states except for the Hardware Layer Failure (defined below)}, as indicated by the dashed arrows with $\lambda_v$ rate in Fig. \ref{mega_seph}. From VLF state, a single transition to the initial state is assumed, because a repairing action of the hypervisor is typically concluded by a complete element restoration. Such an operation usually requires a different activity (with mean duration $1/\mu_{v}$). 
	{	\item The state \textit{Hardware Layer Failure} (HLF) pertains to the hardware failure condition compelling the collapse of hypervisor and all running instances, simultaneously. The corresponding performance vector is $(0,0,...,0)$, as well. Also the HLF state can be reached by any other state, as pointed by the dashed arrows with $\lambda_h$ rate in Fig. \ref{mega_seph}. As in the previous case, from HLF state, a single transition to the initial state is supposed, because a repairing action of hardware layer is presumed to be concluded by a complete VNF rehabilitation. Such an operation requires a technical activity with mean duration $1/\mu_{h}$. }
\end{itemize}
		

{It is useful to clarify that, according to the model in Fig. 3, the failure of the hypervisor or the hardware has effects only on the instances running on the considered physical host, without compromising instances of other VNFs running on different physical devices, possibly located in different geographical regions.}

Each state of the CTMC corresponds to a $K$-dimensional \textit{performance vector} containing the total VNF serving capacity to all tenants. Thus, the \textit{performance level} $g_{i,j}$ offered by the VNF for tenant $i$ in state $j$ can be defined as
\begin{equation}
g_{i,j} = \gamma \cdot \alpha_{i,j}.
\label{eq:gperf}
\end{equation}  

The set containing all possible performance levels of a single VNF is
\begin{equation}
\mathcal{G}=\{\bm{g}_{1}, \bm{g}_{2}, ..., \bm{g}_{N}\},
\label{eq:gvec}
\end{equation}  
where $\bm{g}_{j} = \left(g_{1,j},...,g_{K,j} \right)=\gamma$ $\cdot $ \boldsymbol{$\alpha$}$_j$  is the performance vector in state $j$, and $N$ is the number of states given by
\begin{equation}
N = \prod_{i = 1}^K\left(n_i + 1\right) + 2.
\label{eq:numstati}
\end{equation}

Therefore, the VNF performance level at time $t\geq0$ is modeled by the vector stochastic process $\bm{G}(t) = \left(G_1(t),...,G_K(t)\right) \in\mathcal{G}$ with (state) probability vector $\bm{p}(t)= \left(p_1(t),...,p_N(t)\right)$ at time $t$, where $p_{j}(t) = \mathrm{Pr}\{\bm{G}(t) = \bm{g}_{j}\}$, for $j=1,...,N$. Given an initial probability vector a time $t=0$, $\bm{p}(t)$ is derived at $t\geq0$ by solving the system of differential equations \cite{TrivediBook}
\[
\frac{{\rm d}\bm{p}(t)}{{\rm d}t} = \bm{p}(t) \mathbf{Q}
\]
together with the normalization condition $\sum_{j=1}^N p_j(t) = 1$, where $\mathbf{Q}$ is the infinitesimal generator matrix \cite{rubino1989} of the CTMC shown in Fig. \ref{mega_seph}. Being the performance model $\bm{G}(t) $ an ergodic CTMC, a unique steady-state probability distribution $\bm{p} = \left(p_1,...,p_N\right)$ is the solution of $\bm{p} \; \mathbf{Q}=\bm{0}$, where  
\begin{equation}
p_j = \lim_{t\rightarrow\infty} \mathrm{Pr}\{\bm{G}(t) = \bm{g}_j\},  \;\;\; j = 1, ..., N.
\label{eq:pi}
\end{equation}

Thus, the discrete random vector $\bm{G} = \left(G_1,...,G_K\right)$, representing the asymptotic behavior of $\bm{G}(t)$ as $t\rightarrow\infty$, has values in the set (\ref{eq:gvec}) with probabilities (\ref{eq:pi}). 
The collection of pairs $\left\{p_j, \bm{g}_{j} \right\}$, $j = 1,...,N$, completely determines the steady-state performance behavior (in terms of serving capacity) of a VNF.

\subsection{Availability of the SFC}
\label{sec:VIMS}

\begin{figure}[t!]
	\centering
	\captionsetup{justification=centering}
	\includegraphics[scale=0.45]{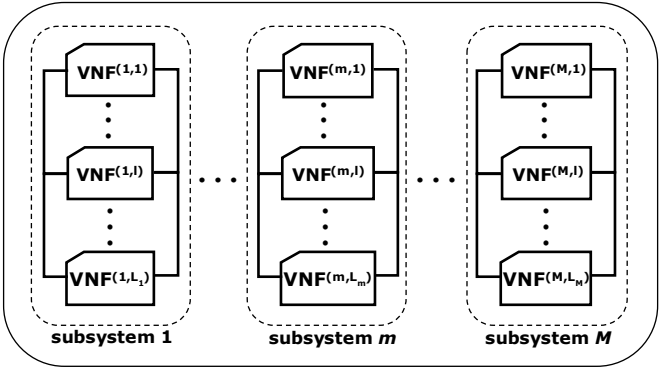}
	\caption{A multi-tenant SFC infrastructure where parallel configuration for each VNF element is considered. VNF$^{(m,l)}$ refers to parallel node $l$ ($l=1,\dots,L_m$) of subsystem $m$ ($m=1,\dots,M$).}
	\label{fig:parallel}
\end{figure} 

An availability model for an SFC is built by considering that: {\it i)} the SFC works when all VNFs are operative (see Fig. \ref{fig:vnffg}); thus, a \textit{series} connection among them is in force; {\it ii)} in order to cope with faults and to guarantee high availability requirements, some redundancy must be introduced. We adopt \textit{parallel} redundancy for each VNF, where some load balancing solutions are also assumed to exploit simultaneously all the parallel VNFs.
The final series/parallel availability model for an SFC is shown in Fig. \ref{fig:parallel}, where VNF$^{(m,l)}$ is the parallel node $l$ of VNF $m$ composing the SFC system, and the  \textit{subsystem} $m$ is the parallel redundant configuration of VNF $m$. 
A multi-tenant SFC system is supposed to be {\em available} when each tenant provides to its customers a required performance level (also referred to as {\em demand}): this calls for the introduction of a $K$-dimensional demand vector $\bm{W}(t)=\left( W_1(t),...,W_K(t) \right)$. To cope with a given demand in the presence of random failures via parallel redundancy, the whole
SFC can be conveniently modeled as a series-parallel system with {\it {flow} dispersion} \cite{Levitin2003}, indicating that any of the parallel elements is able to handle service requests. Therefore, the performance levels of a subsystem composed by parallel VNFs is given by the sum of the performance vectors provided by a single VNF. Finally, the series connection among redundant functionalities imposes that the SFC system performance be limited by the subsystem with the lowest performance level for each tenant. Focusing on tenant $i$, the performance level at time $t$ of the series-parallel SFC system is:
\begin{equation}
G_i^{S}(t)=\min_{m\in\{ \mathrm{1, \dots, \textit{M}}\}}\sum_{l=1}^{L_m}G^{(m,l)}_i(t),
\label{eq:gtot}
\end{equation}
where $G^{(m,l)}_i(t)$ represents the performance level of tenant $i$ exhibited by parallel element $l$ of subsystem $m$ at time $t$, and $L_m$ is the number of parallel nodes of subsystem $m$.

Let $\bm{G}^{S}(t) = \left(G_1^{S}(t),..., G_K^{S}(t)\right)$ be the SFC system performance process, namely the vector stochastic process collecting all tenants performance (\ref{eq:gtot}), $\forall i=1,...,K$. Accordingly, $\bm{G}^{S}(t)$ can be represented for $t\rightarrow\infty$ by a (discrete) random vector $\bm{G}^{S} = \left(G_1^{S},\dots,G_K^{S}\right)$ with a number of (vector) states equal to
\beq
J = \prod_{m=1}^{M} J^{(m)},
\label{eq:VIMSstates}
\eeq
where $J^{(m)} = \prod_{l=1}^{L_m}N^{(m,l)}$ is the number of states of the subsystem $m$, and $N^{(m,l)}$ is given by (\ref{eq:numstati}) for each parallel element $l$ of $m$, where $n_i^{(m,l)}$ software instances are considered for tenant $i$.

Let $\bm{g}^{S}_{j} = \left(g^{S}_{1,j},...,g^{S}_{K,j} \right)$ be the performance level vector $j$ of the SFC, and $p^{S}_{j} = \lim_{t\rightarrow\infty} \mathrm{Pr}\{\bm{G}^{S}(t) = \bm{g}^{S}_{j}\}$ its limiting probability as $t\rightarrow\infty$. 
The collection of pairs 
\beq
\left\{p^{S}_{j}, \bm{g}^{S}_{j} \right\}, j=1,\dots,J
\label{eq:pairsVIMS}
\eeq 
completely determines the steady-state performance behavior of the whole SFC system. 
It is worth noting that a direct solution of the CTMC representing the whole MSS of the SFC system is unfeasible, due to the huge state space cardinality $J$ in (\ref{eq:VIMSstates}). However, hierarchical approaches, like that proposed in Section \ref{sec:ugf}, can help to reduce the required computational burden in finding the steady-state performance distribution of the overall SFC and, then, its steady-state availability. 
This latter can be determined by introducing the {\em instantaneous availability} \cite{Levitin2003} $A^{S}\left[t,\bm{W}(t)\right]$, namely the probability that the system at $t>0$ is in one of the {\em acceptable states}, defined as the states where performance of tenant $i$ is not less than demand $W_i(t)$ for each $i=1,...,K$, viz.
\begin{equation}
A^{S}\left[t,\bm{W}(t)\right]=\mathrm{Pr}\{G_i(t) - W_i(t)\geq 0, \;\; \forall i=1,...,K \}. 
\label{eq:at} 
\end{equation}
As {\em t} grows, the system initial state has a vanishing influence on its availability. Therefore, given a constant demand level $\bm{W}(t) = \bm{w}=\left(w_1,...,w_K\right)$, the {\em steady-state availability} $A^{S}\left(\bm{w}\right) = \lim_{t\rightarrow\infty} A^{S}\left(t,\bm{w}\right)$ can be computed by means of the pairs (\ref{eq:pairsVIMS}) as
\setlength{\arraycolsep}{0.0em}
\begin{eqnarray}
A^{S}(\bm{w}) &{}={}& \displaystyle\sum_{j=1}^{J} p^{S}_{j}\cdot
\bm{1}\left(g^{S}_{i,j}\geq w_i, \forall i = 1,...,K\right),
\label{eq:astat2}
\end{eqnarray}
\setlength{\arraycolsep}{5pt}

\noindent where $\bm{1}(\mathcal{C})=1$ if condition $\mathcal{C}$ holds true and $0$ otherwise.

\section{Availability evaluation of a multi-tenant SFC system}
\label{sec:ugf}

A convenient procedure to compute the steady-state distribution (hence the availability or other dependability metrics of complex MSS systems) is based on the UGF method. It is a hierarchical approach that avoids handling the huge overall state-space model of a complex system (which likely turns out insoluble in most practical configurations) by combining the distributions of its components through some suitable operators amenable to numerical implementations with affordable complexity \cite{Levitin2003}. In case of series-parallel systems, series and parallel operators are needed.
For the sake of clarity, we start by presenting the UGF of an SFC system used by a single tenant ($K = 1$), where scalar performance levels are involved. Then, we generalize the approach to $K > 1$ by introducing the Multidimensional UGF.

\subsection{UGF for single-tenant SFC}
\label{sec:ugfIMS}

The UGF of the (steady-state) performance distribution $G^{(m,l)}$ of parallel node $l$ of subsystem $m$ is the polynomial-shape function (also referred to as $u$-function) defined as:
\beq
u^{(m,l)}(z) = \sum_{j=1}^{N^{(m,l)}}p_j^{(m,l)} z^{g_j^{(m,l)}},
\label{eq:simpleUGF}
\eeq
where ${g_j}^{(m,l)}$ is the performance level in state $j$, and ${p_j}^{(m,l)}$ is the corresponding steady-state probability.

In order to characterize the $u$-function of a system described by a series/parallel availability model, two operators can be adopted: an operator $\pi$ to build the $u$-function of subsystems connected in parallel, and an operator $\sigma$ to calculate the $u$-function of elements interconnected in series. Recall that, under the assumption of adopting load balancing techniques for parallel nodes \cite{Guida2013}, a subsystem constituted by multi-state components with flow dispersion is obtained. According to \cite{globecom10}, the $u$-function of subsystem $m$ with $L_m$ parallel nodes is expressed by the $\pi$ operator that represents the product of the UGFs for each node, namely
\setlength{\arraycolsep}{0.0em} 
\begin{eqnarray} \label{eq:paralleldef}
u^{(m)}(z) &=& \pi \left( u^{(m,1)}(z),\dots, u^{\left(m,L_m\right)}(z) \right) \nonumber \\
&=& \displaystyle\prod_{l=1}^{L_m}\left( \sum_{j_l=1}^{N^{(m,l)}} p^{(m,l)}_{j_l} z^{g^{(m,l)}_{j_l}}\right) \nonumber \\
&=& \displaystyle\sum_{j_1=1}^{N^{(m,1)}}\cdot\cdot\cdot\sum_{j_{L_m}=1}^{N^{\left(m,L_m\right)}}
\left[\left( \prod_{l=1}^{L_m} p^{(m,l)}_{j_l} \right) z^{\sum_{l=1}^{L_m}g^{(m,l)}_{j_l}}\right] \nonumber \\
&=& \displaystyle\sum_{j=1}^{J^{(m)}} p_j^{(m)} z^{g_j^{(m)}}. 
\end{eqnarray}
\setlength{\arraycolsep}{5pt}

On the other hand, the UGF of a series configuration is defined (for a couple of subsystems) as:

\setlength{\arraycolsep}{0.0em}
\begin{align}
\sigma\left( u^{(1)}(z),u^{(2)}(z) \right) =  \displaystyle\sum_{j=1}^{J^{(1)}} \sum_{h=1}^{J^{(2)}} p_{j}^{(1)} p_{h}^{(2)} z^{\min\left\{g_{j}^{(1)},g_{h}^{(2)}\right\}},
\label{eq:simpleseries}
\end{align}
\setlength{\arraycolsep}{5pt}

\noindent where the minimum arises since the element exhibiting the lowest performance level represents the bottleneck in a series-connected system. By applying (\ref{eq:simpleseries}) recursively, the UGF of $M$ subsystems composing the SFC system in Fig. \ref{fig:parallel} is easily derived as: 
\beq
u^{S}(z) = \sigma\left( u^{(1)}(z),u^{(2)}(z),..., u^{(M)}(z)\right), 
\label{eq:sigmadef}
\eeq
whose UGFs are provided by (\ref{eq:paralleldef}), that can be recast as 
\beq
u^{S}(z) = \sum_{j=1}^{J}p^{S}_j z ^{g^{S}_j}, 
\label{eq:SFCsystem}
\eeq
where $J$ is given by (\ref{eq:VIMSstates}).

Through (\ref{eq:SFCsystem}), the performance levels $g^{S}_j$ and the steady-state probabilities $p^{S}_j$ are simply read out as the exponents of $z$ and the respective coefficients. The steady-state availability $A^{S}(w)$ is then obtained by (\ref{eq:astat2}), particularized with $K = 1$ and scalar demand $w$.

\subsection{The Multidimensional UGF}
\label{sec:mugf}

As said before, the availability evaluation of a multi-tenant SFC signaling system involves vector random processes describing performance of all tenants using the system. In order to preserve the benefits of the UGF approach, we propose the {\it Multidimensional} UGF (MUGF) that accounts for the case $K>1$, useful to handle performance vectors such as $\bm{G}^{(m,l)}$ and $\bm{G}$. Indeed, the MUGF approach allows to efficiently combine the collection of pairs  $\left\{p^{(m,l)}_j, \bm{g}^{(m,l)}_j \right\}$ $\forall m,l$, arising from the model in Section \ref{sec:mss}, toward assessing the steady-state performance and availability of the whole SFC.
The main concept underlying our proposal is a ``dimension-wise" extension of the 
$\pi$ and $\sigma$ operators.

Accordingly, in keeping with the definition of multivariate probability generating function, we define the MUGF $u(\bm{z})$ of the $K$-dimensional random vector $\bm{G}$, with values in the set $\left\{\bm{g}_{1},...,\bm{g}_{J}\right\}$  and probabilities in the set $\left\{p_{1},...,p_{J}\right\}$, as
\begin{equation}
u(\bm{z})=\sum_{j=1}^{J} {p_{j} \prod_{i=1}^{K} {z_i^{g_{i,j}}}},
\label{eq:multiu}
\end{equation}
where $\bm{g}_{j} = \left(g_{1,j},...,g_{K,j} \right)$ and $\bm{z} = \left(z_1,...,z_K\right)$.

As a result, the MUGF $u^{(m)}(\bm{z})$ of a subsystem $m$, composed by $L_m$ parallel VNFs \textit{with flow dispersion} \footnote {A ``parallel" version of the MUGF operator has also been proposed for the case of parallel systems {\em without flow dispersion} \cite{dimauro2017}.}, amounts to the following extension of (\ref{eq:paralleldef}):

\setlength{\arraycolsep}{0.0em}
\begin{eqnarray}\label{eq:mugfpi}
u^{(m)}(\bm{z}) &=& \pi \left( u^{(m,1)}(\bm{z}),\dots,u^{\left(m,L_m\right)}(\bm{z}) \right)  \nonumber \\
&=& \displaystyle\prod_{l=1}^{L_m}\left[\sum_{j_l = 1}^{N^{(m,l)}} \left( p_{j_l}^{(m,l)} \prod_{i=1}^{K}z_i^{g^{(m,l)}_{i,j_l}} \right)\right]   \nonumber\\
&=& \displaystyle\sum_{j_1=1}^{N^{(m,1)}}\cdots\sum_{j_{L_m} = 1}^{N^{\left(m,L_m\right)}} \left(\prod_{l=1}^{L_m} p_{j_l}^{(m,l)} \right) \prod_{i=1}^{K}z_i^{\sum_{l=1}^{L_m}g^{(m,l)}_{i,j_l}} \nonumber\\
&=& \displaystyle\sum_{j=1}^{J^{(m)}} p_j^{(m)} \prod_{i=1}^{K}z_i^{g_{i,j}^{(m)}},
\label{eq:16}
\end{eqnarray}
\setlength{\arraycolsep}{5pt}

\noindent where  $u^{(m,1)}(\bm{z}),\dots,u^{\left(m,L_m\right)}(\bm{z})$ represent the MUGFs of nodes composing subsystem $m$ with $J^{(m)}$ different states, characterized by performance levels vectors $\bm{g}^{(m)}_{j}=\left(g_{1,j}^{(m)},...,g_{K,j}^{(m)}\right)$ and occurrence probability $p_j^{(m)}$.

Once drawn $u^{(m)}(\bm{z})$ by (\ref{eq:mugfpi}), the MUGF $u^{S}(\bm{z})$ of a multi-tenant SFC system is computed by the following series operator
\beq
u^{S}(\bm{z}) = \sigma\left( u^{(1)}(\bm{z}),u^{(2)}(\bm{z}),...,u^{(M)}(\bm{z}) \right),
\label{eq:mugfsigma}
\eeq
which can be elaborated by applying recursively the $K$-dimensional version of the binary operator (\ref{eq:simpleseries}), viz.:
\setlength{\arraycolsep}{0.0em}
\begin{eqnarray}
&\sigma& \left( u^{(1)}(\bm{z}),u^{(2)}(\bm{z}) \right) \nonumber \\
&{}={}& \displaystyle\sum_{j=1}^{J^{(1)}} \sum_{h=1}^{J^{(2)}} p_{j}^{(1)} p_{h}^{(2)} \prod_{i=1}^{K}z_i^{\min\left\{g_{i,j}^{(1)},g_{i,h}^{(2)}\right\}},
\label{eq:mugfseries}
\end{eqnarray}
\setlength{\arraycolsep}{5pt}

\noindent where $J^{(1)}$ ($J^{(2)}$) is the number of performance vectors $\bm{g}_{j}$ ($\bm{g}_{h}$), whose probability is $p_{j}^{(1)}$ ($p_{h}^{(2)}$), of subsystem $m_1$ ($m_2$). 

Finally, $u^{S}(\bm{z})$ can be recast, like (\ref{eq:SFCsystem}), as
\beq
u^{S}(\bm{z}) =  \displaystyle\sum_{j=1}^{J}p^{S}_j \prod_{i=1}^{K}z_i^{g^{S}_{i,j}},
\label{eq:MSFCsystem}
\eeq
where $J$ is given by (\ref{eq:VIMSstates}).
Therefore, $u^{S}(\bm{z})$ is a polynomial-shape function in $K$ indeterminates $z_1,\dots, z_K$, where each term provides the performance vector $\bm{g}^{S}_{j} = \left(g_{1,j}^{S},...,g_{K,j}^{S}\right)$ (exponents of $z_i$), and its steady-state probability $p^{S}_j$ (corresponding coefficient).  It is worth noting that the sum in (\ref{eq:MSFCsystem}) collects all the terms with the same exponents $\bm{g}^{S}_{i,j}$ (by summing the corresponding probabilities), thus the resulting number of the  effective performance levels can be much less than $J$.

The steady-state availability $A^{S}(\bm{w})$ of the multi-tenant SFC system is finally provided by (\ref{eq:astat2}).

Once granted an expression for $A^{S}(\bm{w})$, it becomes possible \cite{ushakov1987} to address some redundancy optimization problem by exhaustive search or other known techniques, such as genetic algorithms \cite{Levitin2003,levitin98}.  

Letting $C^{(m,l)}$ be the cost of parallel node $l$ in subsystem $m$, the overall cost of the multi-tenant SFC configuration  {$\bm{l}=\left(L_1,...,L_M\right)$} is
\beq
C^{S}(\bm{l})=\sum_{m=1}^{M} \sum_{l = 1}^{L_m} C^{(m,l)}.
\label{eq:cost}
\eeq
%
A problem of interest when designing a multi-tenant SFC service is to devise the configuration $\bm{l}^*$ that minimizes the total cost of deployment while satisfying a certain steady-state availability requirement $A_0$. 
Given the set $\mathcal{L}^{S} = \left\{ \bm{l}:  A^{{S}}(\bm{w},\bm{l})\geq A_0 \right\}$ of the possible configurations satisfying the steady-state availability condition, the formal solution of the problem amounts to:
\beq
\bm{l}^*= \argmin_{\bm{l} \in \mathcal{L}^{S} } C^{S}(\bm{l}) .
\label{eq:optprob}
\eeq

\section{A numerical example}
\label{sec:results}

In this section, we provide an example of availability analysis focusing on a cutting-edge deployment of a virtualized SFC infrastructure: the IP Multimedia Subsystem \cite{eri2014}, \cite{Garcia2016}. 
IMS has been elected by the telecommunication industry as the enabling technology of 5G networks, providing a huge variety of IP-based services ranging from real-time multimedia (i.e. phone calls) to web messaging.
The IMS signaling network functionalities are called Call Session Control Functions (CSCFs), and are distributed among three servers that communicate mainly by exchanging Session Initiation Protocol (SIP) messages: the \textit{Proxy} CSCF (PCSCF), acting as an interface between a user and the IMS network; the \textit{Serving} CSCF (SCSCF), performing some core functions such as session and routing control or user registration management; the \textit{Interrogating} CSCF (ICSCF), forwarding SIP requests or responses to the appropriate SCSCF.
Another key element is the \textit{Home Subscriber Server} (HSS), an advanced database containing users' profiles that can be queried by means of Diameter, a specific protocol.

By exploiting the SFC paradigm, IMS can be deployed as a service chain named virtualized IMS (vIMS).

\begin{figure}[t!]
	\centering
	\mbox{\subfloat[]{\includegraphics[scale=0.44]{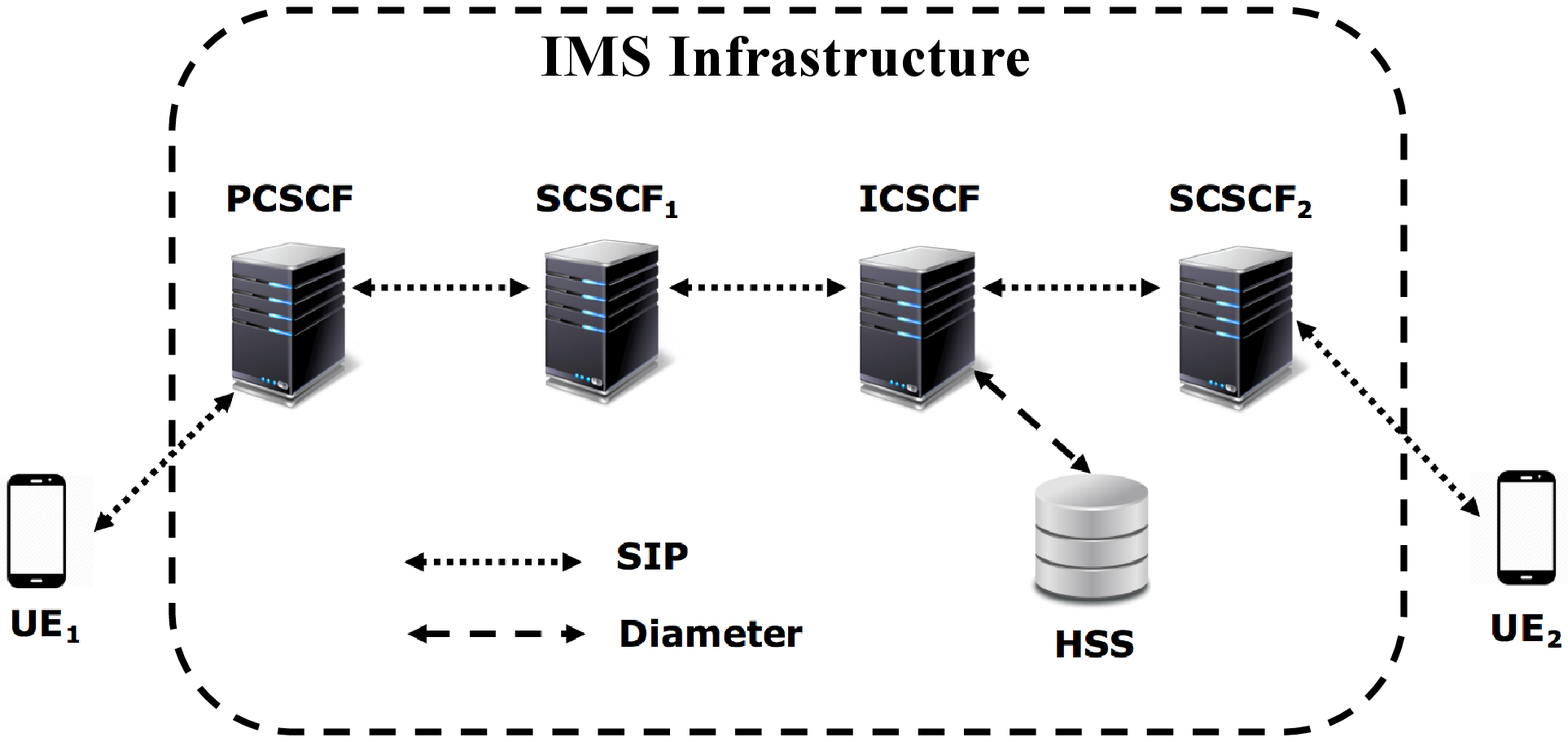}}}
	\mbox{\subfloat[]{\includegraphics[scale=0.29, angle=90]{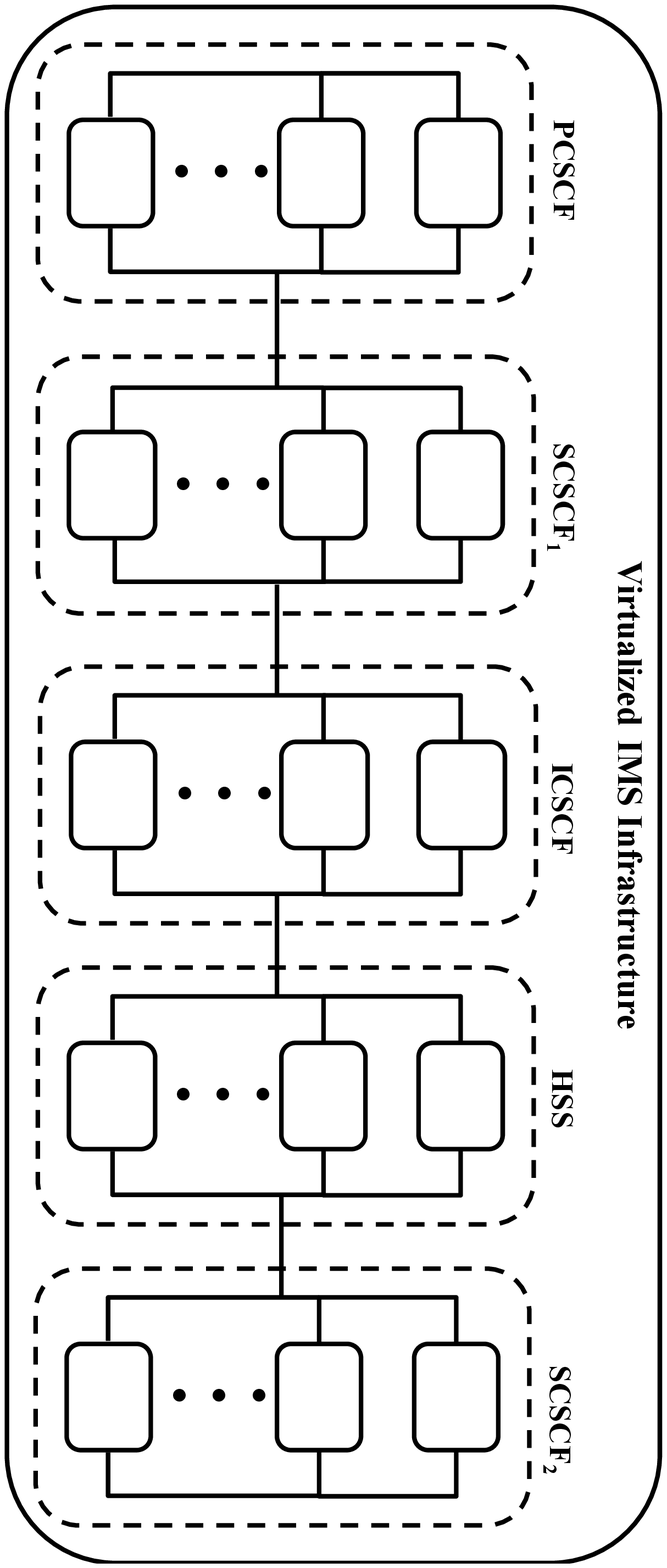}}}
	\caption{(a) The IMS signaling network for call set-up establishment between two UEs. (b) A multi-tenant vIMS representation, where parallel redundancy for each IMS server is considered.} 
	\label{megafig}
\end{figure}

In this example, we focus on the call set-up procedure between two mobile phones.
The considered scenario is represented in Fig. \ref{megafig}(a), where  the signaling flow, originated by a calling User Equipment (UE\textsubscript{1}), traverses the IMS servers in an ordered way to reach a called UE (UE\textsubscript{2}). In particular, the SIP request of UE\textsubscript{1} is forwarded by the server PCSCF (the first contact point of IMS network) towards SCSCF\textsubscript{1}. In order to reach UE\textsubscript{2}, SCSCF\textsubscript{1} forwards such a request to server I that, after querying HSS, can identify server SCSCF\textsubscript{2} in charge of managing the network area where UE\textsubscript{2} is located. Upon completion of the call set-up procedure, UE\textsubscript{1} and UE\textsubscript{2} can establish a multimedia session, e.g. an audio/video call.

Figure \ref{megafig}(b) shows the same system in a multi-tenant SFC environment, where a chain of parallel vIMS elements (all shared among the various operators) has been introduced in order to achieve the required redundancy for high availability of the call set-up service.

To exemplify the MUGF approach introduced in Section \ref{sec:ugf}, {we compute the steady-state availability of a minimal deployment cost of the vIMS system, by solving the redundancy optimization problem (\ref{eq:optprob}) with a given (steady-state) availability requirement.}

We assume for simplicity that the nodes composing the vIMS series-parallel system have one and the same performance model like that proposed in Section \ref{sec:mss}, where relevant parameters are: number of states $N^{(m,l)} = N$; {failure and repair rates for the hardware layer $\lambda_{h}^{(m,l)} =\lambda_{h}$ and $\mu_{h}^{(m,l)} =\mu_{h}$, respectively;} failure and repair rates for the virtualization layer $\lambda_{v}^{(m,l)} =\lambda_{v}$ and $\mu_{v}^{(m,l)} =\mu_{v}$, respectively; failure and repair rates for the software instances $\lambda_{s_i}^{(m,l)} = \lambda_{s_i}$, $\mu_{s_i}^{(m,l)} =\mu_{s_i}$, respectively; number of the software instances implementing the service logic for a given tenant $i$ $n_i^{(m,l)} = n_i$, $\forall i=1,...,K$. We assume also that all nodes have one and the same cost, that we suppose, with no lack of generality, equal to $1$, i.e. $C^{(m,l)} = 1$, $\forall l \in \left\{1,...,L_m\right\}$, with $m$ indexing the set $\left\{ \mathrm{PCSCF},\mathrm{SCSCF_1},\mathrm{ICSCF},\mathrm{HSS},\mathrm{SCSCF_2} \right\}$. However, these assumptions can be easily generalized to more realistic scenarios, with different performance characteristics (due to services offered by IMS servers) and costs imposed by different manufacturers. 

\begin{figure}[t!]
	\centering
	\captionsetup{justification=centering}
	\includegraphics[scale=0.38,angle=90]{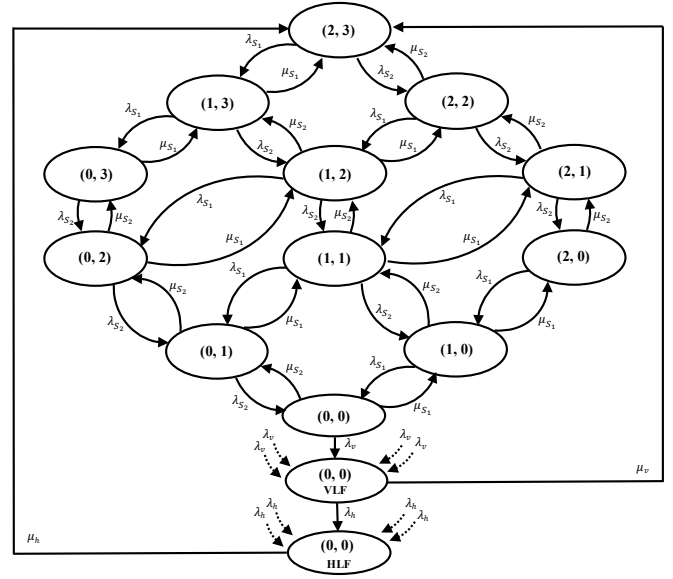}
	\caption{A multi-state model of a virtualized IMS node (VNF) with 2 tenants; the tenant 1 is supposed to manage 2 software instances and the tenant 2 is supposed to manage 3 software instances.}
	\label{fig:miniseph}
\end{figure}

\begin{figure*}[t!]
	\begin{align}
	\centering
	\captionsetup{justification=centering}
	\includegraphics[scale=0.72]{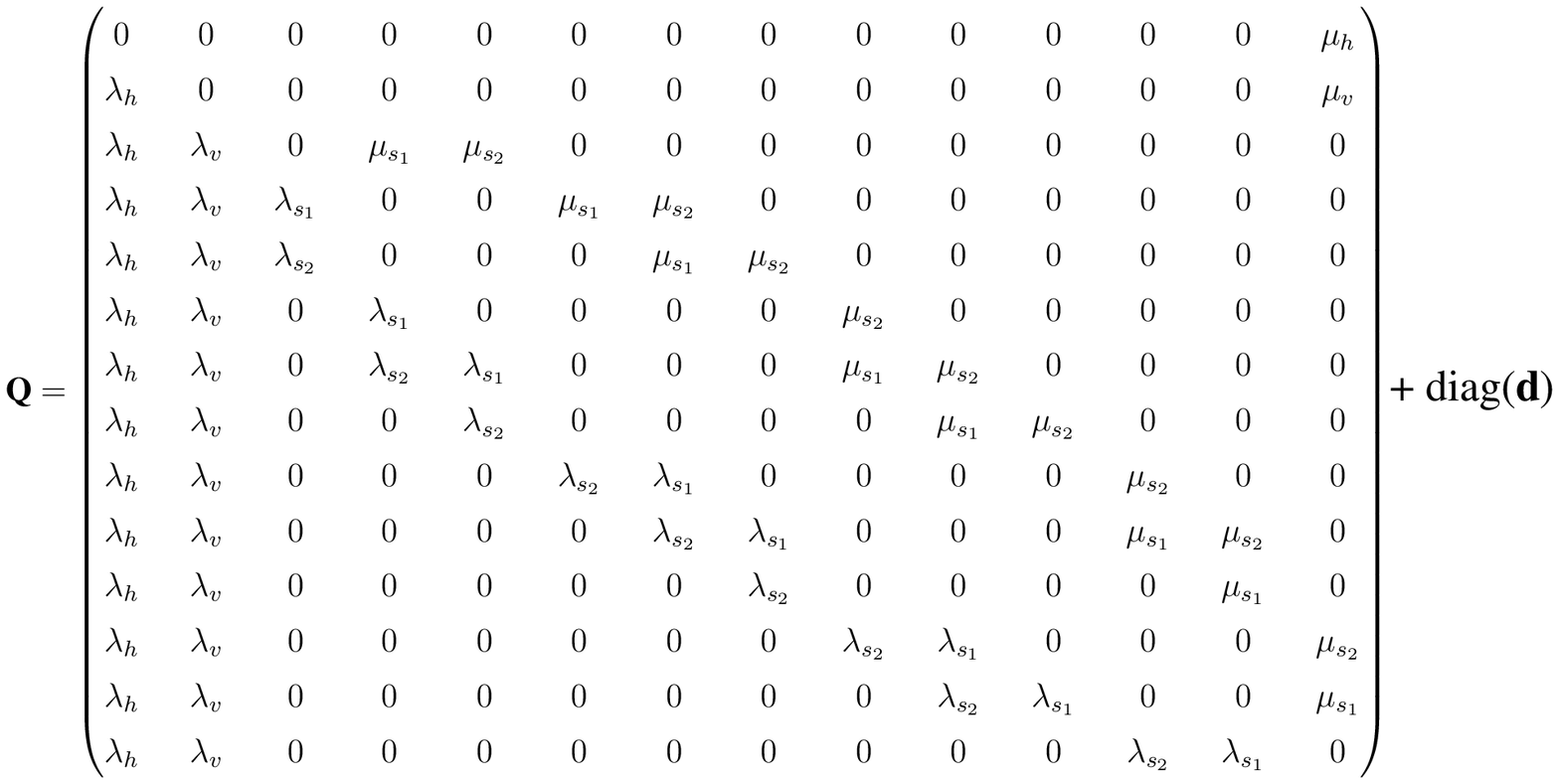}
	\label{eq:infgen}
	\end{align}
\end{figure*}

In our exemplary scenario, $K=2$ telecom operators (tenants) are using the vIMS system and offer different service levels to their customers. Precisely, we assume a (constant) demand vector $\bm{w} = (w_1, w_2) = (15000, 25000)$ sessions.
Given the serving capacity $\gamma = 10000$ sessions, we choose, somewhat arbitrarily, $n_1 = \lceil{w_1 / \gamma}\rceil = 2$ and $n_2 = \lceil{w_2 / \gamma}\rceil = 3$, as these are the smallest integers allowing to satisfy the demand without redundancy and without faults.
The resulting MSS performance model of a vIMS node (thus representing a VNF) is a special case of the CTMC depicted in Fig. \ref{mega_seph}, with {$N = (n_1+1)(n_2+1) + 2 = 14$} different states (after applying (\ref{eq:numstati})) described by a $2$-dimensional performance vector containing the number of the active software instances for each tenant.{Such an exemplary MSS is depicted in Fig. \ref{fig:miniseph}, where, in accordance with the model presented in Fig. \ref{mega_seph}, a fully working system is embodied in the state $(2,3)$, whereas, a failed system is simultaneously accounted by states $(0,0)$, VLF and HLF}.
Failure and repair rates of software instances are assumed the same for both tenants and in line with telecommunication experts' hints and with pertinent literature \cite{trivedi2012}: $\lambda_{s_1} = \lambda_{s_2} = 1.587\times 10^{-6}$ s$^{-1}$ (corresponding to 1 fault every 175 hours) and $\mu_{s_1} = \mu_{s_2} = 5.556\times 10^{-4}$ s$^{-1}$ (corresponding to a mean repair time of 30 minutes). Failure and repair rates of the virtualization part are $\lambda_{v} = 1.047\times 10^{-7}$  s$^{-1}$ (corresponding to 1 fault every 2654 hours) and $\mu_{v} = 1.667\times 10^{-4}$  s$^{-1}$ (corresponding to a mean repair time of 100 minutes), respectively.
{Finally, failure and repair rates of the hardware layer are $\lambda_{h} = 4.630\times 10^{-9}$  s$^{-1}$ (corresponding to 1 fault every 60000 hours) and $\mu_{h} = 3.472\times 10^{-5}$  s$^{-1}$ (corresponding to a mean repair time of 8 hours), respectively.}
By solving this CTMC with $N=14$ states for $t\rightarrow\infty$ as described in Sect. \ref{sec:mss}, the steady-state performance distribution (in terms of the number of call set-up sessions) for a single virtualized node is computed, and is given by the collection of pairs $\left\{p_j, \bm{g}_{j} \right\}$, where $p_j$ is the probability (\ref{eq:pi}) correlated to the performance level $\bm{g}_{j}$ in the set (\ref{eq:gvec}) and where $\mathbf{Q}$ is given by (\ref{eq:infgen}), along with the vector \textbf{d} defined as:
\begin{align*}
\textbf{d}=\left(-\mu_h, {-}\lambda_h {-} \mu_v, -\lambda_h {-} \lambda_v {-} \sum \mu_s, -\lambda_v {-} \lambda_h {-}  \lambda_{s_1} \right. \\ \nonumber
 \left.    {-}  \sum \mu_s, -\lambda_h {-} \lambda_v {-} \lambda_{s_2} {-}  \sum \mu_s, {-} \lambda_h {-} \lambda_v  {-} \lambda_{s_1} {-} \mu_{s_2}, \right. \\ \nonumber 
 \left.   {-} \lambda_h {-} \lambda_v {-} \sum \lambda_s {-} \sum \mu_s, {-} \lambda_h {-} \lambda_v {-} \lambda_{s_2} {-} \sum \mu_s, \right. \\ \nonumber
\left.    {-} \lambda_h {-} \lambda_v {-} \sum \lambda_s {-}  \mu_{s_2}, {-} \lambda_h {-} \lambda_v {-} \sum \lambda_s {-} \sum \mu_s, \right. \\ \nonumber
\left.   {-} \lambda_h {-} \lambda_v {-}  \lambda_{s_2} {-} \mu_{s_1}, -\lambda_h {-} \lambda_v {-}  \sum \lambda_s {-} \mu_{s_2},  \right. \\  \nonumber
\left.   -\lambda_h {-} \lambda_v {-}  \sum \lambda_s {-} \mu_{s_1}, \lambda_{s_1}, -\lambda_h {-} \lambda_v {-}  \sum \lambda_s \right). \nonumber
\end{align*}
Table \ref{tab:perfnode} summarizes the obtained performance levels and pertinent state probabilities. It is useful to note that $p_1+p_2+p_3$ refers to $(0,0)$ performance vector since HLF, VLF and the state accounting for all failed software instances admit the same performance vector (namely, a completely failed system). 

The corresponding MUGF $u(\bm{z})  = u(z_1,z_2)$ is reported in (\ref{eq:singlenodemugf}).

\begin{table}[t]
	\renewcommand{\arraystretch}{1.5}
	\caption {{Steady-state probabilities and performance levels in terms of the number of call set-up sessions of a virtualized node.}}
	\label{tab:perfnode}
	\begin{center}   
		\begin{tabular}{c | c | c}
			\hline
			State & Probability & Performance \\ probabilities & values & levels
			\\ 
			\hline\hline
			\\[-8pt]
			$p_1 + p_2 + p_3$ & $7.608\times 10^{-4}$  & $(0,0)$ \\
			$p_4$ & $6.617\times 10^{-11}$ & $(0,10000)$ \\
			$p_5$ & $2.316\times 10^{-8}$ & $(0,20000)$ \\
			$p_6$ & $8.107\times 10^{-6}$  & $(0,30000)$ \\
			$p_7$ & $6.617\times 10^{-11}$ & $(10000,0)$ \\
			$p_8$ & $2.316\times 10^{-8}$ & $(10000,10000)$ \\
			$p_9$ & $8.108\times 10^{-6}$ & $(10000,20000)$ \\
			$p_{10}$ & $2.838\times 10^{-3}$ & $(10000,30000)$ \\
			$p_{11}$ & $2.316\times 10^{-8}$ & $(20000,0)$ \\
			$p_{12}$ & $8.107\times 10^{-6}$  & $(20000,10000)$ \\
			$p_{13}$ & $2.838\times 10^{-3}$ & $(20000,20000)$ \\
			$p_{14}$ & $0.9935$ & $(20000,30000)$ \\
			\hline
		\end{tabular}
	\end{center}
\end{table}

\begin{figure*}[!ht]
	\setlength{\arraycolsep}{0.0em}
	\begin{eqnarray}
	u(\bm{z})  &{}={}&  7.608 \times10^{-4} + 6.617 \times10^{-11} \; {z_{1}}^{10000} + 2.316 \times10^{-8} \; {z_{1}}^{20000} + 6.617 \times10^{-11} \; {z_{2}}^{10000} \nonumber\\
	&&{+}\: 2.316 \times10^{-8} \; {z_{1}}^{10000} \; {z_{2}}^{10000} + 8.107 \times10^{-6} \; {z_{1}}^{20000} \; {z_{2}}^{10000} + 2.316 \times10^{-8} \; {z_{2}}^{20000} \nonumber\\
	&&{+}\: 8.108 \times10^{-6} \; {z_{1}}^{10000} \; {z_{2}}^{20000} + 2.838 \times10^{-3} \; {z_{1}}^{20000} \; {z_{2}}^{20000} + 8.107 \times10^{-6} \; {z_{2}}^{30000}  \nonumber\\
	&&{+}\: 2.838 \times10^{-3} \; {z_{1}}^{10000} \; {z_{2}}^{30000} + 0.9935 \; {z_{1}}^{20000} \; {z_{2}}^{30000}
	\label{eq:singlenodemugf}
	\end{eqnarray}
	\setlength{\arraycolsep}{5pt}
	\hrulefill
\end{figure*}

\begin{figure*}[!t]
	\setlength{\arraycolsep}{0.0em}
	\begin{eqnarray}
	u^{I}(\bm{z}) &{}={}& 5.806 \times10^{-7} + 1.011 \times10^{-13} \; {z_{1}}^{10000} + 3.540 \times10^{-11} \; {z_{1}}^{20000} + 3.149 \times10^{-18} \; {z_{1}}^{30000} \nonumber\\
	&&{+}\: 5.412 \times10^{-16} \; {z_{1}}^{40000} + 1.011 \times10^{-13} \; {z_{2}}^{10000} \; + 3.540 \times10^{-11} \; {z_{1}}^{10000} \; {z_{2}}^{10000} \nonumber\\
	&&{+}\: 1.239 \times10^{-8} \; {z_{1}}^{20000} \; {z_{2}}^{10000} + 2.204 \times10^{-15} \; {z_{1}}^{30000} \; {z_{2}}^{10000} + 3.789 \times10^{-13} \; {z_{1}}^{40000} \; {z_{2}}^{10000} \nonumber\\
	&&{+}\: 3.540 \times10^{-11} \; {z_{2}}^{20000} + 1.239 \times10^{-8} \; {z_{1}}^{10000} \; {z_{2}}^{20000} + 4.338 \times10^{-6} \; {z_{1}}^{20000} \; {z_{2}}^{20000} \nonumber\\
	&&{+}\: 1.157 \times10^{-12} \; {z_{1}}^{30000} \; {z_{2}}^{20000} + 1.990 \times10^{-10} \; {z_{1}}^{40000} \; {z_{2}}^{20000} + 1.239 \times10^{-8} \; {z_{2}}^{30000} \nonumber\\
	&&{+}\: 4.338 \times10^{-6} \; {z_{1}}^{10000} \; {z_{2}}^{30000} + {\underline{1.519 \times10^{-3} \; {z_{1}}^{20000} \; {z_{2}}^{30000}}} + {\underline{5.403 \times10^{-10} \; {z_{1}}^{30000} \; {z_{2}}^{30000}}} \nonumber\\
	&&{+}\: {\underline{9.287 \times10^{-8} \; {z_{1}}^{40000} \; {z_{2}}^{30000}}} + 1.654 \times10^{-15} \; {z_{2}}^{40000} + 1.158 \times10^{-12} \; {z_{1}}^{10000} \; {z_{2}}^{40000} \nonumber\\
	&&{+}\: {\underline{6.079 \times10^{-10} \; {z_{1}}^{20000} \; {z_{2}}^{40000}}} + {\underline{1.418 \times10^{-7} \; {z_{1}}^{30000} \; {z_{2}}^{40000}}} + {\underline{2.438 \times10^{-5} \; {z_{1}}^{40000} \; {z_{2}}^{40000}}} \nonumber\\ 
	&&{+}\: 3.858 \times10^{-13} \; {z_{2}}^{50000} + 2.701 \times10^{-10} \; {z_{1}}^{10000} \; {z_{2}}^{50000} + {\underline{1.418 \times10^{-7} \; {z_{1}}^{20000} \; {z_{2}}^{50000}}} \nonumber\\
	&&{+}\: {\underline{3.310 \times10^{-5} \; {z_{1}}^{30000} \; {z_{2}}^{50000}}} + {\underline{5.690 \times10^{-3} \; {z_{1}}^{40000} \; {z_{2}}^{50000}}} + 6.632 \times10^{-11} \; {z_{2}}^{60000} \nonumber\\
	&&{+}\: 4.643 \times10^{-8} \; {z_{1}}^{10000} \; {z_{2}}^{60000} + {\underline{2.438 \times10^{-5} \; {z_{1}}^{20000} \; {z_{2}}^{60000}}} + {\underline{5.690 \times10^{-3} \; {z_{1}}^{30000} \; {z_{2}}^{60000}}} \nonumber\\
	&&{+}\:{\underline{0.9870 \; {z_{1}}^{40000} \; {z_{2}}^{60000}}}
	\label{eq:supermugf}
	\end{eqnarray}
	\setlength{\arraycolsep}{5pt}
	\hrulefill
\end{figure*}

\setcounter{figure}{7}
\begin{figure}[t!]
	\centering
	\captionsetup{justification=centering}
	\includegraphics[scale=0.46]{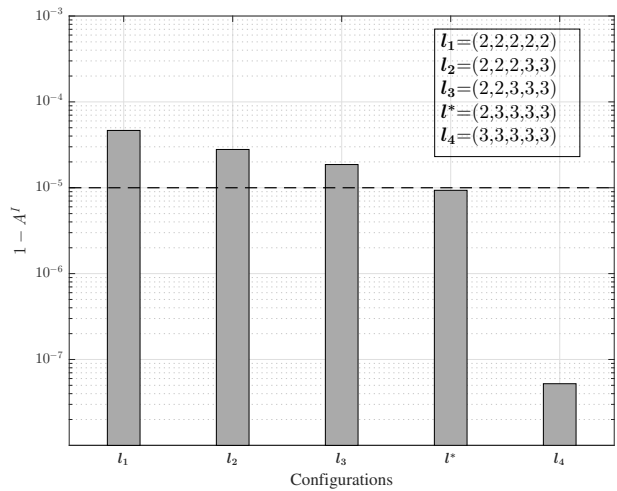}
	\caption{Unavailability $1-A^{I}$ of the virtualized IMS infrastructure with five exemplary redundant configurations $\bm{l}$. The horizontal dashed line is the required steady-state unavailability $1-A^{I}=10^{-5}$. The minimal cost configuration guaranteeing the ``five nines" condition is $\bm{l^*}$.} 
	\label{fig:unavail}
\end{figure} 

\setcounter{figure}{8}
\begin{figure*}[!t] 
	\centering
	\subfloat[]{
		\includegraphics[width=0.44\linewidth]{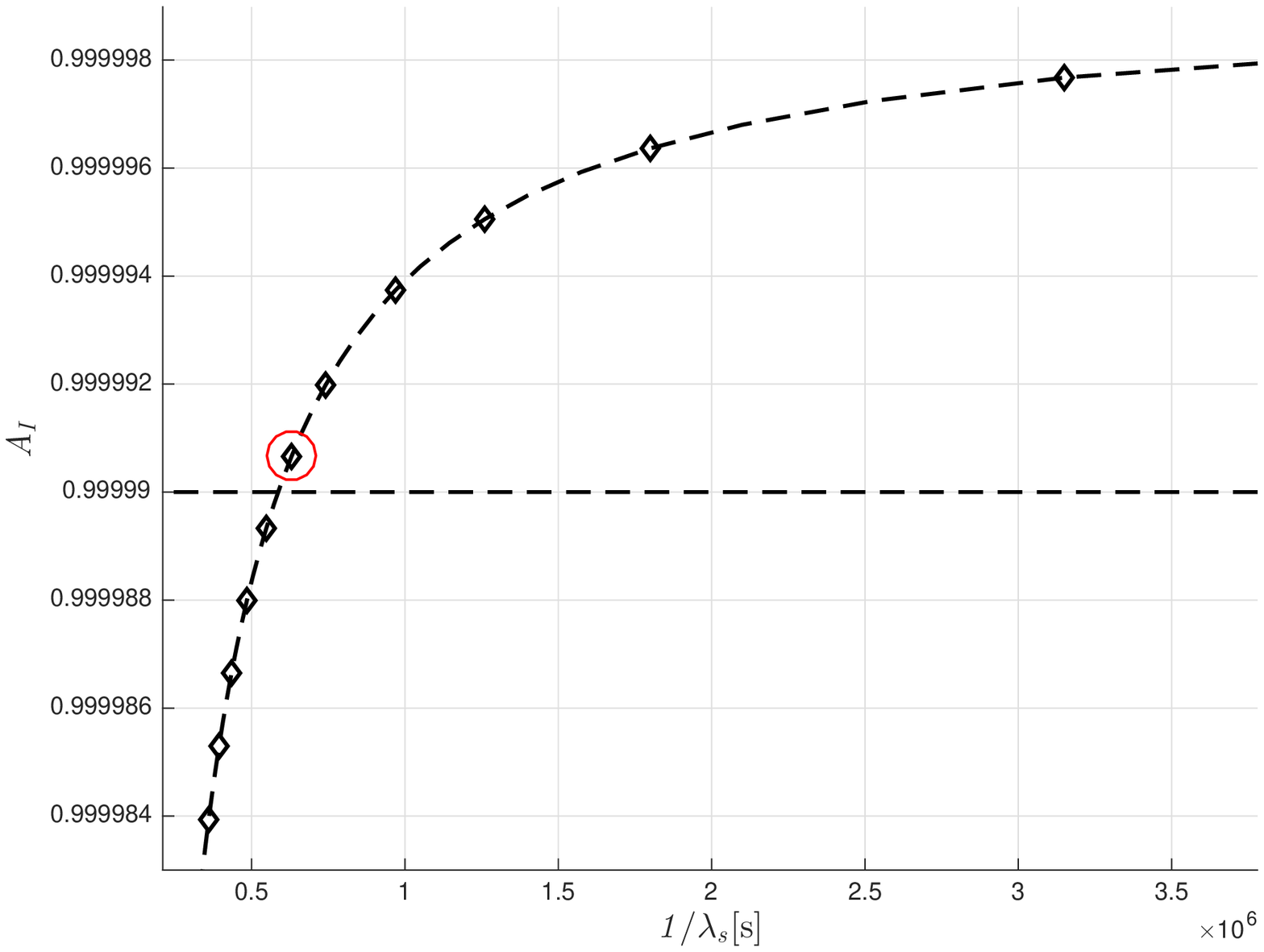}}
	\label{fig:sens_lams} \vspace{1em} \hspace{2em} 
	\subfloat[]{
		\includegraphics[width=0.44\linewidth]{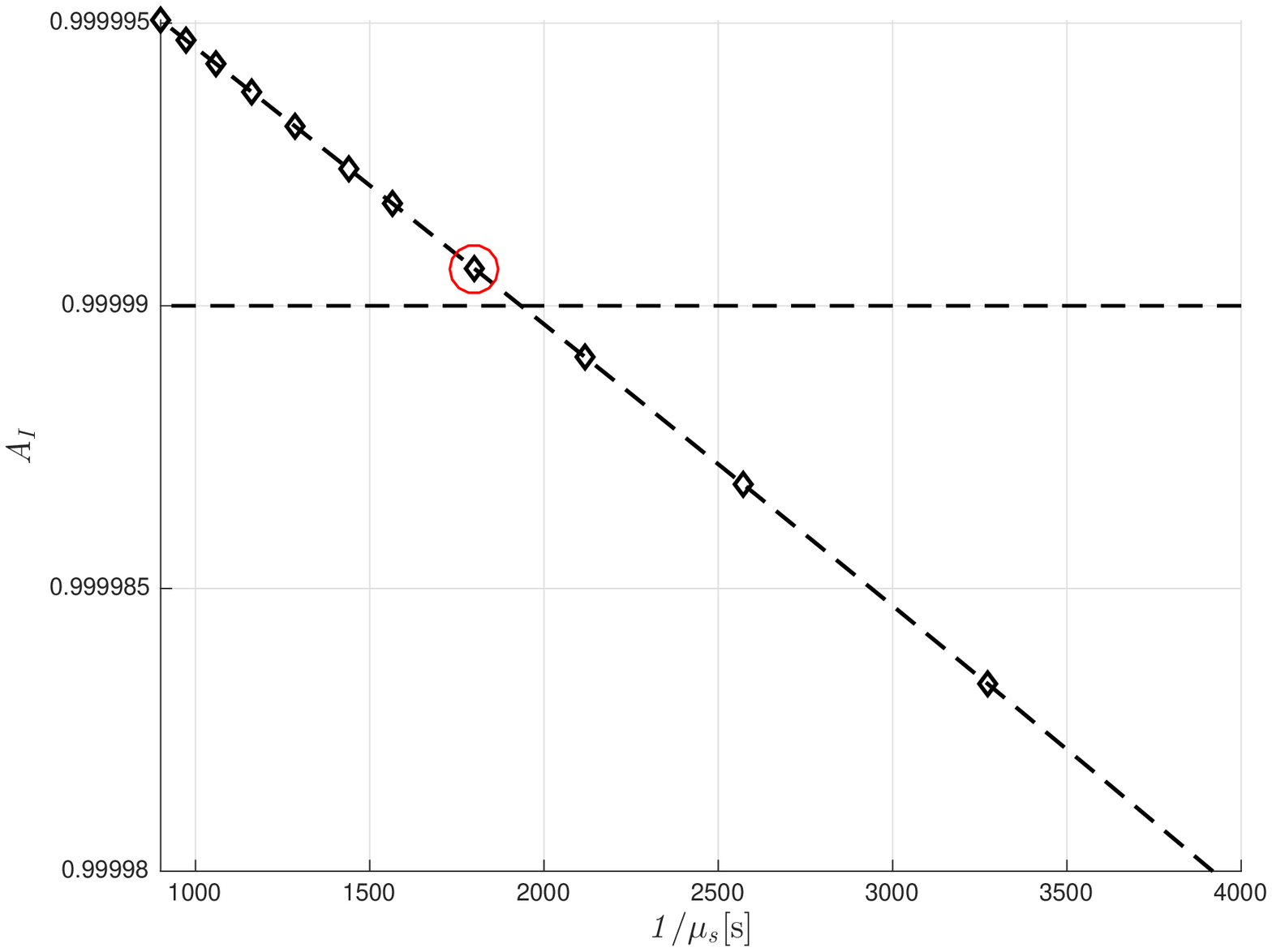}}
	\label{fig:sens_mus}
	\subfloat[]{
		\includegraphics[width=0.44\linewidth]{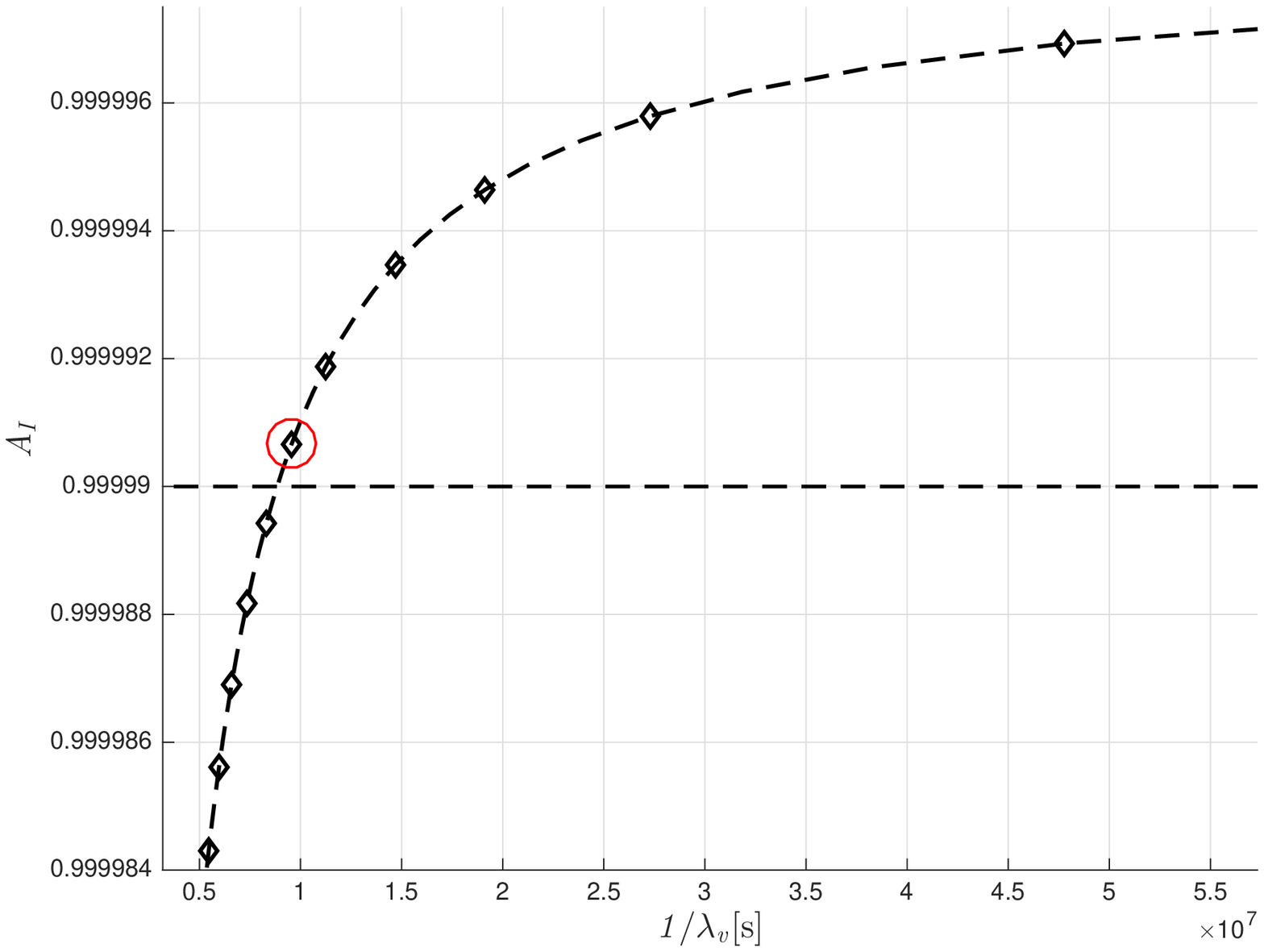}}
	\label{fig:sens_lamv} \hspace{2em}
	\subfloat[]{
		\includegraphics[width=0.44\linewidth]{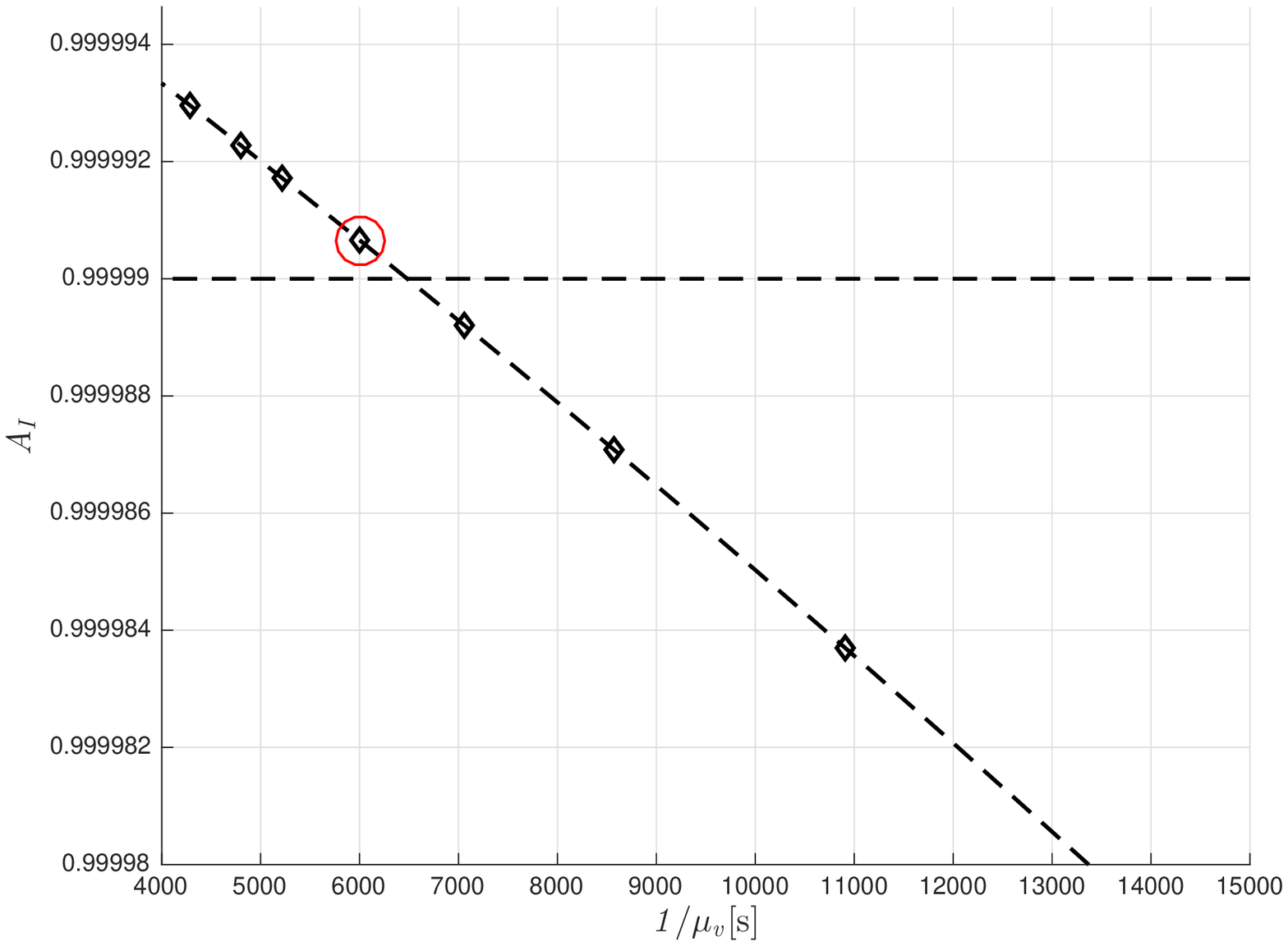}}
	\label{fig:sens_muv}    	
		\subfloat[]{
			\includegraphics[width=0.44\linewidth]{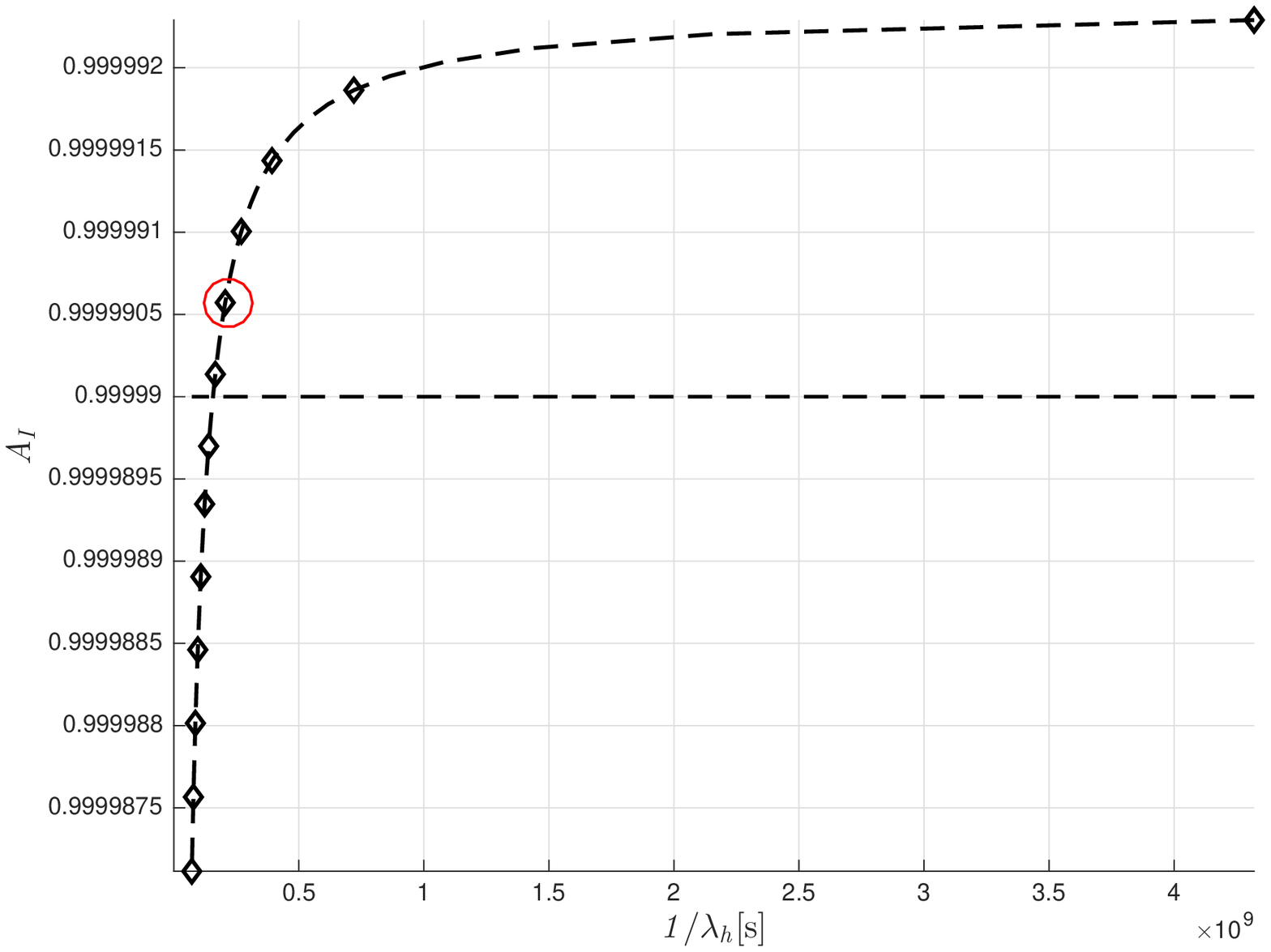}}
		\label{fig:sens_lamh} \hspace{2em}
		\subfloat[]{
			\includegraphics[width=0.44\linewidth]{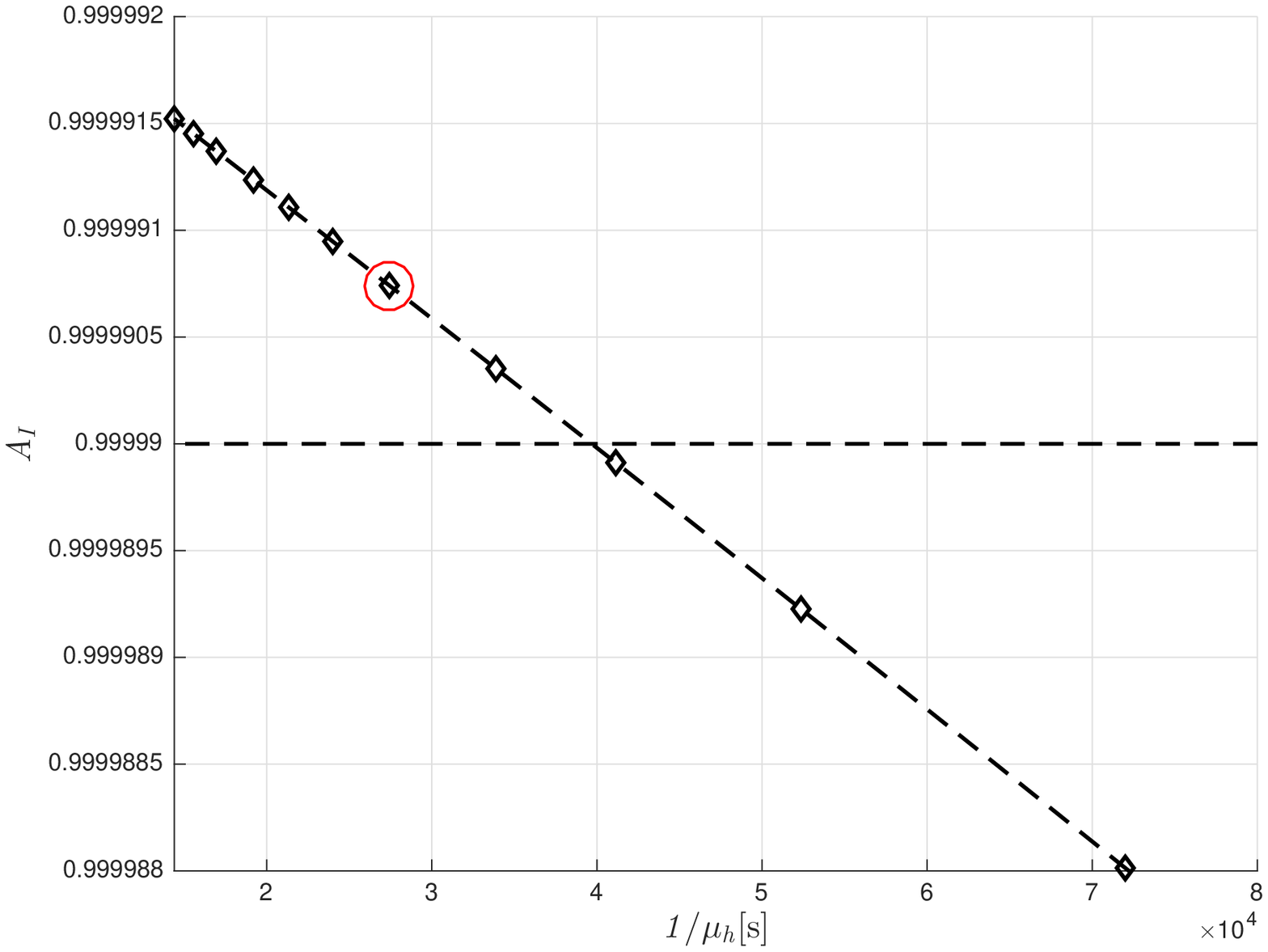}}
		\label{fig:sens_muh}     
	
	\caption{Influence on the overall vIMS infrastructure of: software instances failure rate (a), software instances repair rate (b), virtualization layer failure rate (c), virtualization layer repair rate (d), {hardware layer failure rate (e), hardware layer repair rate (f)}. Nominal values are circled in red.}
	\label{fig:sens_all}
\end{figure*}

In order to meet the ``five nines" availability requirement for the vIMS, we set $A_0 = 1 - 10^{-5}$ and solve numerically the optimization problem (\ref{eq:optprob}) by an exhaustive search approach, having imposed a uniform limitation of $4$ redundant nodes for each server.
A routine (available upon request for non commercial use), written in Mathematica\textsuperscript{\textregistered}, implements the proposed MUGF approach: it evaluates the steady-state availability of the vIMS system (\ref{eq:astat2}) in every redundancy configuration by computing the MUGF (\ref{eq:MSFCsystem}) via parallel and series operators (\ref{eq:mugfpi}) and (\ref{eq:mugfseries}), respectively. 
Then, the numeric algorithm finds the minimal cost configuration(s) $\bm{l^*}$.
In this use case, $5$ (almost) equivalent solutions were found: they correspond to $\bm{l^*}$ consisting of the combinations of $4$ (arbitrarily chosen) subsystems with $3$ redundant nodes, and the remaining subsystem with $2$ redundant nodes (we briefly indicate such a configuration by $\bm{l^*}=(2,3,3,3,3)$). 

The MUGF $u^{I}(\bm{z})$ of $\bm{l^*}$ is reported in (\ref{eq:supermugf}), where the probabilities and the performance levels of the acceptable states (those with performance levels greater than $15000$ and $25000$ for the first and the second tenant, respectively) are underlined and highlighted (in green in the electronic version). The resulting steady-state availability, in terms of the number of call set-up sessions managed by the vIMS system, is computed as the sum of the probabilities of the acceptable states and is equal to $A^{I}(\bm{w},\bm{l}^*) = 0.999993413$ while the cost is $C^{I}(\bm{l}^*)=14$.
An availability analysis performed by considering some alternative redundant configurations $\bm{l}$ of vIMS servers offers the results shown in Fig. \ref{fig:unavail}, where, for a more comfortable visualization, the unavailability $1-A^{I}(\bm{w},\bm{l})$ is reported. The horizontal dashed line represents the target steady-state unavailability $1-A_{0}=10^{-5}$. It is readily seen that configurations $\bm{l_1}=(2,2,2,2,2)$ (with $C^{I}(\bm{l_1})=10$), $\bm{l_2}=(2,2,2,3,3)$ (with $C^{I}(\bm{l_2})=12$) and $\bm{l_3}=(2,2,3,3,3)$ (with $C^{I}(\bm{l_3})=13$) do not meet the availability requirement. 
Remarkably, adding just $1$ extra redundant component, the configuration $\bm{l_4}=(3,3,3,3,3)$ (with $C^{I}(\bm{l_4})=15$) achieves the much larger availability value of $0.9999999646$.  Thus, although exceeding the specifications, configuration $\bm{l_4}$ might be more appealing to the network designer.
We remark that the MUGF approach circumvents the computational burden inherent in the baseline  CTMC approach.  Indeed, considering our example  with $N^{(m,l)}=N=14$, the state space of the CTMC is worth  $J = 14^{\sum_{m=1}^5 L_m} = 14^n$, being $n$ the number of nodes in the system. Thus, the optimization problem (\ref{eq:optprob}) amounts to the solution  of  $4^5$ linear systems of $J$ equations, with $J$ ranging between  $14^5 \cong 5.4\cdot 10^5$ and  $14^{20} \cong 8.4 \cdot 10^{22}$; in particular, for  the optimal configuration $l^* = (2,3,3,3,3)$, $J=14^{\left(\sum_{m=1}^4 3 \right)\left( \sum_{m=1}^1 2\right)} = 14^{14} \cong 1.1 \cdot 10^{16}$. Needless to say, the baseline approach is unfeasible with these numbers. 
On the other hand, the MUGF technique takes advantage of a hierarchical decomposition of the problem as follows: {\it 1)} the steady-state distribution of each component VNF, namely a CTMC with $N = 14$ states, is computed by solving a system of $N$ equations; {\it 2)} the computed distributions are combined via the series/parallel operators, which involve the standard algebraic manipulations in (\ref{eq:16}), (\ref{eq:mugfsigma}) and (\ref{eq:mugfseries}). Coming to our example, step  {\it 1)} of  MUGF approach amounts to the solution of a system of $N = 14$ equations  that requires about $10$ ms for each VNF on a notebook based on an Intel Core i7–4960 HQ CPU@2.60GHz; whereas step {\it 2)} requires, for each configuration, an average time of $73$ ms on the same platform, and hence about $75$ s  to complete the exhaustive search over  $4^5$ redundancy configurations.
After having determined the solution $\bm{l}^*$, we have performed a sensitivity analysis aimed at evaluating the robustness of vIMS system with respect to deviation of some critical parameters from their nominal values. 
In particular, the panel of Figs. \ref{fig:sens_all} shows the influence of failure rates $\lambda_s$, $\lambda_v$, $\lambda_h$, and repair rates $\mu_s$, $\mu_v$, $\mu_h$ (all expressed in terms of their reciprocals) on the overall system availability. 
In each figure, a circle (in red in the electronic version) points out the nominal value of the parameter under analysis, whereas the horizontal dashed line represents the ``five nines" limit value. 

Figures \ref{fig:sens_all}(a), \ref{fig:sens_all}(c), \ref{fig:sens_all}(e) highlight that $\bm{l}^*$ configuration still meets the ``five nines" availability requirement for slightly higher fault rates up to (approximately) $1$ fault every $167$ hours, $1$ fault every $2527$ hours and $1$ fault every $44440$ hours, for software instances, virtualization layer and hardware layer, respectively. 
Similarly, Figs. \ref{fig:sens_all}(b), \ref{fig:sens_all}(d), \ref{fig:sens_all}(f) show that the nominal values $1/\mu_s$, $1/\mu_v$, $1/\mu_h$ can be relaxed up to (approximately) a mean repair time of $31.6$ minutes, $105$ minutes and $10.7$ hours, for software instances, virtualization layer and hardware layer, respectively, still satisfying the ``five nines" condition.

Finally, we have analyzed the effects of variations of  $\bm{w}$ around its nominal value $(15000, 25000)$, by determining the respective optimal configurations $\bm{l}^*$ and the pertinent availability values. Table \ref{tab:variable_demand} reports the results corresponding to increments or decrements to the initial value in blocks of $5000$ for both tenants.  
As expected, the more demanding $\bm{w}$, the greater the number of redundant elements and SFC cost are, and vice versa.

 \begin{table}[t]
 	\fontsize{10}{1}
 	\renewcommand{\arraystretch}{1.9}
 	\caption {Effects of variation of $\bm{w}$.}
 	\label{tab:variable_demand}
 	\begin{center}   
 		\begin{tabular}{c | c | c}
 			\hline
 			$\bm{w}$ &  $\bm{l}^*$ & $A^{I}(\bm{w},\bm{l}^*)$
 			\\ 
 			\hline \hline
 			$(15000, 25000)$ & $(2,3,3,3,3)$  & $ 0.999990659$ \\
 			$(20000, 20000)$ & $(2,2,3,3,3)$  & $0.999990022$ \\
 			$(20000, 30000)$ & $(2,3,3,3,3)$  & $0.999990659$ \\
 			$(10000, 30000)$ & $(2,2,3,3,3)$  & $0.999990114$ \\
 			$(10000, 20000)$ & $(2,2,2,2,2)$  & $0.999996982$ \\
 			\hline
 		\end{tabular}
 	\end{center}
 \end{table}

\section{Conclusions}
\label{sec:conclusions}

Today, the service composition is becoming a common practice for network and telecommunication operators desiring to boost the provisioning of novel services. In this spirit, the Service Function Chaining (SFC), supported by virtualization concepts introduced by Network Function Virtualization (NFV), proposes an infrastructure built on virtualized network functions (or VNFs) to be traversed in an ordered way aimed at providing specific services. In many applications, the VNFs belonging to the service chain are intended to host many instances of different operators (or tenants), resulting in a multi-tenant environment.  

In this paper we have afforded an availability analysis of a multi-tenant SFC infrastructure by offering a threefold contribution. First, we have modeled a multi-tenant SFC infrastructure as a multi-state system by conveniently combining series and parallel operators. Then, we have proposed an extended version of the Universal Generating Function (UGF) technique, referred to as Multidimensional UGF (MUGF), useful to cope with performance vectors applicable to complex multi-tenant network scenarios. Finally, we have performed an availability analysis in a realistic scenario of a virtualized IP Multimedia Subsystem (vIMS), a state-of-the-art deployment of a multi-tenant SFC infrastructure. The vIMS steady-state availability has been computed by choosing the number of call set-up requests (handled by the system for each operator with different performance requirements) as performance vector. Accordingly, given a service demand vector and a high system availability target, a parallel redundancy optimization problem has also been solved by a computationally efficient routine implementing MUGF approach, and the network configurations minimizing the cost expressed in terms of number of deployed nodes have been identified.
Future work will be devoted to include in the model more sophisticated dependencies among nodes behavior, typically present in real world scenarios.

\vspace{-30pt}
\begin{IEEEbiography}[{\includegraphics[width=1in,height=1.15in,clip,keepaspectratio]{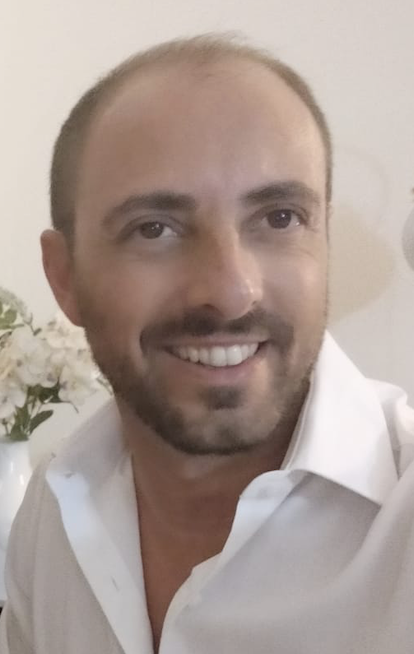}}] 
{Mario Di Mauro} received the Laurea degree in electronic engineering from the University of Salerno (Italy) in 2005, the M.S. degree in networking from the University of L'Aquila (Italy) jointly with the Telecom Italia Centre in 2006, and the PhD. degree in information engineering in 2018 from University of Salerno.
He was a Research Engineer with CoRiTel (Research Consortium on Telecommunications, led by Ericsson Italy) and then a Research Fellow with University of Salerno. He has authored several scientific papers, and holds a patent on a telecommunication aid for impaired people. His main fields of interest include: network availability, network security, data analysis for
telecommunication infrastructures.
\end{IEEEbiography}

\vspace{-30 pt}

\begin{IEEEbiography}[{\includegraphics[width=1in,height=1.15in,clip,keepaspectratio]{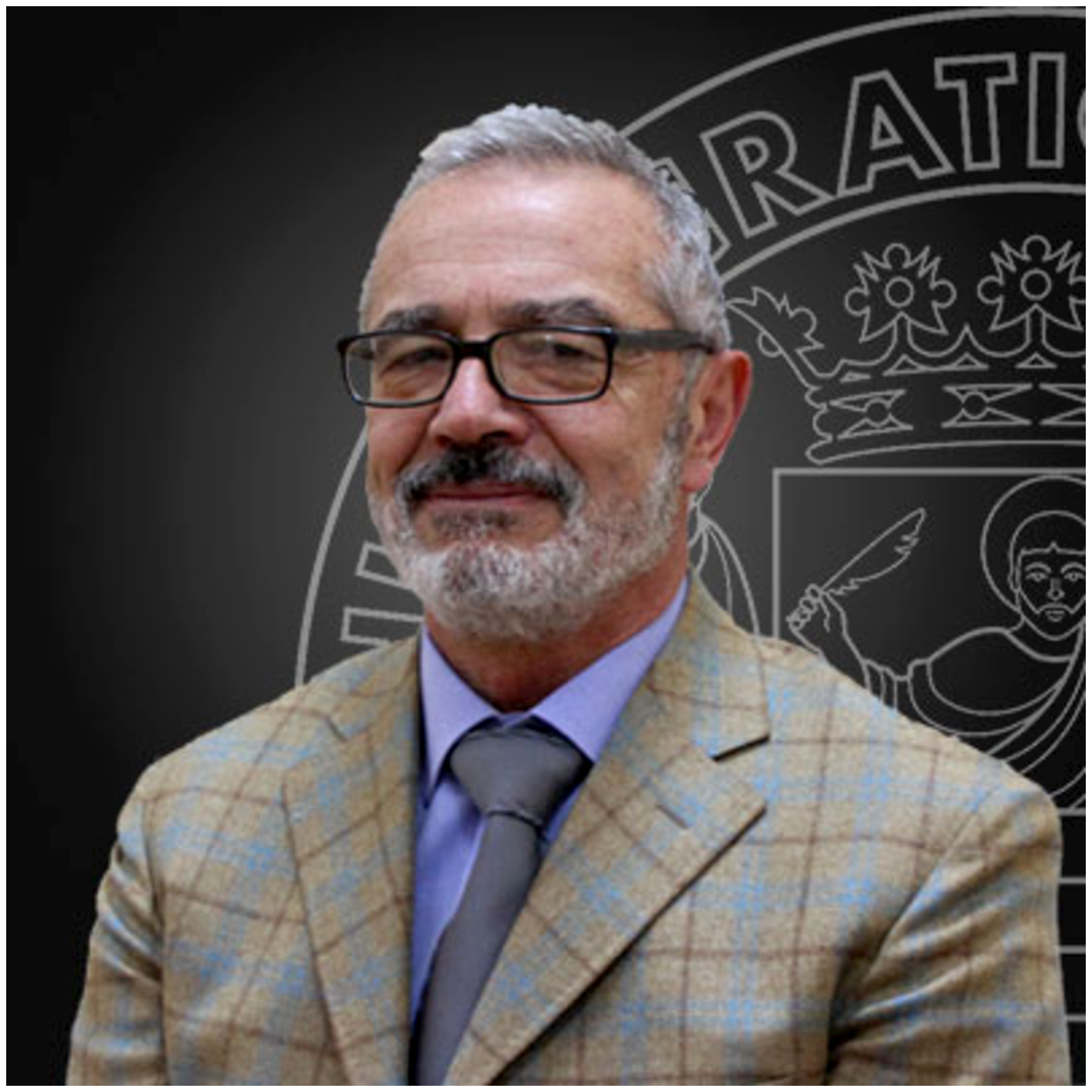}}]
{Maurizio Longo} received the MSEE degree from Stanford University, CA, USA, in 1977, and the Laurea degree in electronic engineering from the University of Napoli (Italy), in 1972. He is currently with the University of Salerno (Italy) as Full Professor of Telecommunications and the Director of the CoRiTel (Research Consortium
on Telecommunications) Lab., having also served as the Department Dean and the 
Chairman of the Graduate School of Information Engineering.
He held academic positions also with the University Federico II (Napoli), the Parthenope University (Napoli), the University of Lecce and the Aeronautical Academy. In 1986 - 1987 and 1990, he was on leave with Stanford University, as a Formez Fellow and as a NATO-CNR Senior Fellow. He has authored over 180 papers in international journals and conference proceedings, mainly in the fields of telecommunication networks and statistical signal processing.
\end{IEEEbiography}
\vspace{-30 pt}
\begin{IEEEbiography}[{\includegraphics[width=1in,height=1.15in,clip,keepaspectratio]{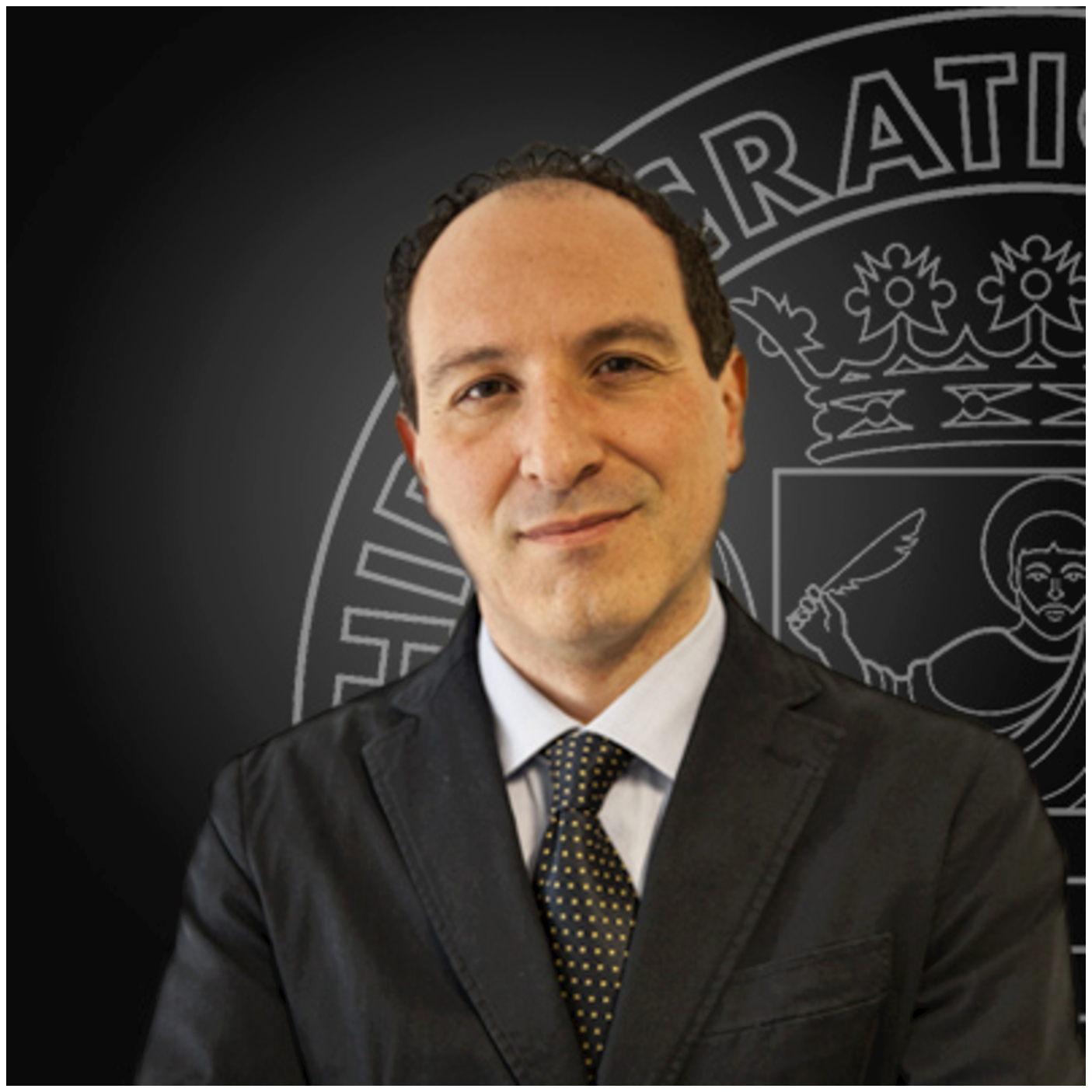}}]
{Fabio Postiglione} is currently an Assistant Professor of Applied Statistics with the Dept. of Information and Electrical Engineering and Applied Mathematics (DIEM) at University of Salerno (Italy). He received his Laurea degree (summa cum laude) in Electrical Engineering and his Ph.D. degree in Information Engineering from University of Salerno in 1999 and 2005, respectively. His main research interests include degradation analysis, lifetime estimation, reliability and availability evaluation of complex systems (telecommunication networks, fuel cells), Bayesian statistics and data analysis. He is/was involved in several EU-funded FP7/H2020 research projects on degradation analysis, lifetime estimation and diagnosis of fuel cells. He is member of the LIGO-VIRGO Collaboration devoted to gravitational waves observation on topics related to data analysis. He has authored over 100 papers, mainly published in international journals.
\end{IEEEbiography}


\end{document}